\documentclass[useAMS,usenatbib,babel]{mnras}

\usepackage[english,english]{babel}
\usepackage[normalem]{ulem}
\usepackage{amsmath}
\usepackage{amssymb,amsfonts,textcomp}
\usepackage{array}
\usepackage{supertabular}
\usepackage{hhline}
\usepackage{hyperref}
\usepackage[squaren, Gray, cdot]{SIunits}
\usepackage[usenames]{color}
\usepackage{graphicx}
\usepackage{caption}
\usepackage{setspace}
\usepackage{multirow,balance}
\usepackage{todonotes}
\usepackage{soul}
\usepackage{float}
\usepackage{times}

\definecolor{lblue}{rgb}{0.1,0.7,1.}
\definecolor{Orange}{rgb}{1.0,0.05,0.15}
\definecolor{Green}{rgb}{0.15,0.45,0.25}
\definecolor{lGreen}{rgb}{50,205,50}
\definecolor{Blue}{rgb}{0.0,0.08,0.65}
\definecolor{Brown}{rgb}{0.7,0.25,0.0}
\definecolor{Pink}{rgb}{1.0,0.05,0.5}
\definecolor{linkcol}{rgb}{0.79, 0.0, 0.09}
\definecolor{grey}{rgb}{0.75,0.75,0.75}
\definecolor{darkerred}{rgb}{0.8,0,0}
\definecolor{darkerblue}{rgb}{0,0,0.8}
\definecolor{Red}{rgb}{0.65,0.08,0.05}

\hypersetup{colorlinks, linkcolor=linkcol, citecolor=blue, filecolor=black, urlcolor=black}

\def\gtrsim{\lower.5ex\hbox{$\; \buildrel > \over \sim \;$}}

\def \apjl {{\it ApJ Let.}}
\def \apjs {{\it ApJ Sup.}}

\def \aap {{\it A\&A}}

\begin{document}

\author[Song, Laigle et al.]{
\parbox[t]{\textwidth}{Hyunmi Song\thanks{E-mail: hmsong@yonsei.ac.kr}$^{1,2}$, Clotilde Laigle$^{3}$, Ho Seong Hwang$^{2}$,
Julien Devriendt$^{4}$,
Yohan Dubois$^{3}$,
Katarina Kraljic$^{5}$,
Christophe Pichon$^{3,6}$,
Adrianne Slyz$^{4}$,
Rory Smith$^{2}$}
\vspace*{6pt} \\ 
$^{1}$ Department of Astronomy, Yonsei University, 50 Yonsei-ro, Seodaemun-gu, Seoul 03722, Republic of Korea\\
$^{2}$ Korea Astronomy and Space Science Institute (KASI), 776 Daedeokdae-ro, Yuseong-gu, Daejeon 34055, Republic of Korea\\
$^{3}$ Sorbonne Universit{\'e}s, UPMC Univ Paris 6 et CNRS, UMR 7095, Institut d'Astrophysique de Paris, 98 bis bd Arago, 75014 Paris\\
$^{4}$ Sub-department of Astrophysics, University of Oxford, Keble Road, Oxford OX1 3RH\\
$^{5}$ Aix Marseille Universit\'e, CNRS, CNES, Laboratoire d'Astrophysique de Marseille, Marseille, France\\
$^{6}$ Korea Institute for Advanced Study (KIAS), 85 Hoegiro, Dongdaemun-gu, Seoul, 02455, Republic of Korea
}
\date{Accepted . Received ; in original form }

\title[The role of cosmic filaments on galaxy mass assembly]
{Beyond halo mass: quenching galaxy mass assembly at the edge of filaments} 
\maketitle

\begin{abstract}
We examine how the mass assembly of central galaxies depends on their location in the cosmic web.
The {\sc Horizon-AGN} simulation is analysed at $z\sim2$ using the {\sc DisPerSE} code to extract multi-scale cosmic filaments.
We find that the dependency of galaxy properties on large-scale environment is mostly inherited from the (large-scale) environmental dependency of their host halo mass. 
When adopting a residual analysis that removes the host halo mass effect, we detect a direct and non-negligible influence of cosmic filaments. 
Proximity to filaments enhances the build-up of stellar mass, a result in agreement with previous studies.
However, our multi-scale analysis also reveals that, at the edge of filaments, star formation is suppressed.
In addition, we find clues for compaction of the stellar distribution at close proximity to filaments.
We suggest that gas transfer from the outside to the inside of the haloes (where galaxies reside) becomes less efficient closer to filaments, due to high angular momentum supply at the vorticity-rich edge of filaments.
This quenching mechanism may partly explain the larger fraction of passive galaxies in filaments, as inferred from observations at lower redshifts.
\end{abstract}


\section{Introduction}
In a now well-established feature of our galaxy formation model, the galaxy stellar mass assembly primarily depends on the mass assembly of their host dark matter haloes. 
One of the most straightforward consequence of this dependency is the stellar-to-halo mass relationship, which has been now extensively studied up to high redshift \citep[e.g.,][and references therein]{zaritsky94,bullock02,berlind02,cooray02,behroozi10,hwang2016,legrand19}. 
At first order, the efficiency of stellar mass assembly is therefore a function of halo mass, owing to the varying modes of gas accretion (cold or hot) and feedback processes (from stars, supernovae and AGN) depending on the depth of the gravitational potential. 
This halo-dependent growth drives differences between the stellar and halo mass functions at the low- and high-mass ends \citep[e.g.,][]{dekel86,bower12,mosteretal10,mosteretal13,lu14,Eckert16,lim17,behroozi19}, as well as different clustering properties of galaxies depending on their type and masses \citep[e.g.,][]{arnouts02,coupon12,bethermin14,hatfield16}. 
One and two-point correlation functions are therefore somehow the most basic statistics used to constrain the galaxy evolution model.  

Environment modulates halo growth, and therefore the growth of galaxies therein.
Although quite generic and sometimes ambiguous, the term `environment' generally refers primarily to the distribution of the matter density field at a scale larger than the halo. 
Large-scale overdensities in the matter density field enable the dark matter proto-halo to pass earlier the critical threshold to collapse, inducing an overabundance of massive haloes as a function of the \textit{amplitude} of the density \citep{bondetal1991}, a trend introduced observationally as the mass-density relation \citep{oemler74}, and further explored in numerous works \citep[e.g.,][]{baldry06,cooper07,hwang19}.

In addition, if the halo is not isolated but resides in the vicinity of another structure (either another halo, or a not fully collapsed overdensity, such as a cosmic wall or a filament), tides may need to be accounted for to model its growth.
This tidal field is shaped by the \textit{geometry} of the matter distribution in the halo's neighbourhood and can induce delays in the accretion rate onto haloes, resulting in e.g., more numerous small, old and stalled haloes in the vicinity of the most massive structures \citep[e.g.][]{shethettormen2004,Hahnetal2009}. 
\cite{Mussoetal2017} showed that, as a signature of the tides from the cosmic web, accretion rate onto haloes is expected to be in general larger at the nodes than in filaments, and also larger in filaments than further away \textit{at fixed mass and density}.
\cite{Hahnetal2009}, and later \cite{borzyskowski17} and \cite{Paranjape18} further showed that the tidal environment of haloes allows them to be separated into two distinct populations.
The small haloes embedded in large-scale filaments undergo tidal suppression of their accretion rate, and therefore quenching of their mass assembly at a level which depends on the amplitude of the anisotropic velocity shear.
In contrast the most massive haloes tend to gravitationally dominate their surroundings, and therefore live in a rather isotropic environment, which will favour their growth.
Hence not only the proximity of the halo to its nearest filament, but also its relative mass with respect to the filament in which it is embedded is a key quantity to understand its long-term evolution. 

In parallel to these theoretical developments, as observational surveys have grown deeper, better sampled and generally more precise, cosmic web extraction from such surveys has become within reach, making possible the quantification of galaxy growth in the natural metric of the cosmic web.
Most of these studies reveal that, at fixed density, galaxies that are closer to cosmic filaments tend to be more massive, quenched, and deficient in HI \citep[see e.g.,][]{alpaslanetal2016, Malavasi2017,chenetal2017, poudeletal2017, laigle18, odekon18,  kraljic18, Sarronetal2019, bonjeanetal2019, salerno19, bird2020}. 
Some works argue however that density is the only driver of these changes \citep[e.g.,][]{Gohetal2019}, and some others claim even opposite trends, in particular regarding the HI gas content \citep[e.g.,][]{kleiner17}. 
The disparity of these results is probably due to the difficulty of properly estimating the density, and due to the  different scales involved \citep[see e.g., a discussion about the different scales of filaments and tendrils in][]{odekon18}, together with mass ranges and redshift ranges which are considered. 

From the observational side, the difficulty however is mostly to disentangle the respective roles of the environment on the one hand and halo mass on the other hand in driving these trends.
The measured dependency of dark matter halo mass and accretion rate on their large-scale environment \citep[e.g.,][]{hahnetal2007,borzyskowski17,Mussoetal2017,Paranjape18} could in fact explain most of the dependency of galaxy mass assembly on large-scale environment, without advocating any additional process specific to baryons.
Due to the difficulty of estimating individual halo masses from observational surveys, it is not clear though from the previously mentioned studies if the environmental dependency of galaxy properties is a consequence of the environmental dependency of halo properties, or if environment impacts baryonic and dark matter differently, at fixed halo mass.

Correctly modelling galaxy formation in the cosmic web is also still a numerical challenge because it requires resolving baryonic physics within galaxies, while correctly sampling their cosmic web environment over statistically significant cosmological volumes. 
In other words, we need to follow trillions of baryons and dark matter particles down to parsec-level scales over Gpc$^{3}$ boxes. 
Current state-of-the art hydrodynamical simulations, such as {\sc Horizon-AGN} \citep{duboisetal14}, Illustris \citep{vogelsbergeretal14}, Eagle \citep{schayeetal15}, Magneticum \citep{dolagetal17}, IllustrisTNG \citep{TNG2018}, and Simba \citep{dave2019}, although impressive, are still limited to box sizes of at most a couple of hundreds Mpc a side, a volume still quite small to make statistically robust statements on the dependency of galaxies on their large-scale environment \citep[although some larger promising simulations are emerging, see e.g.,][for the Horizon Run 5 simulation]{Leeetal2020}.
Furthermore, their maximal resolution usually is about a fraction of a kpc, i.e., too coarse to properly resolve the flows of cold gas penetrating dark matter haloes and feeding galaxies as well as the details of galaxy morphologies, especially at low mass.

However, we still can address the following question: does a cosmological simulation which broadly reproduces galaxy mass assembly exhibit a correlation between the cosmic web environment and galaxy properties beyond halo mass and local density?
In this work, we will use the {\sc Horizon-AGN} simulation which has been shown to correctly reproduce the observed galaxy growth \citep{Kaviraj16} and AGN quenching efficiency \citep{beckmann17}. 
We will take particular care in disentangling the impact of the environment at three different scales: the dark matter halo, the local density and the cosmic filaments.
By doing so, we will examine the unique role of each environmental factor on galaxy mass assembly.
We will emphasize in particular the role played by cosmic filaments in driving additional diversity in galaxy properties (e.g., the scatter of the stellar to halo mass relation) beyond what is driven by halo mass or local density. 
Previous environmental analysis on {\sc Horizon-AGN} have focused on the alignment of galaxies with cosmic filaments \citep{duboisetal14,welkeretal14,welker18,welkeretal20,codis18,kraljic19a}, the connectivity of galaxies and galaxy groups \citep{darraghford19,kraljic19b}, or the evolution of galaxy properties in the frame of the saddle \citep{kraljic18}.
The analysis that we propose in this paper is complementary to the one presented in the latter, but explores smaller scales with a different approach. 

Section~\ref{Sec:pres} presents the {\sc Horizon-AGN} simulation, the halo and galaxy catalogues, and the environment estimators. 
Properties of haloes and galaxies are measured at $z=1.97$.
Section~\ref{Sec:methodology} presents our method of analysis. Section~\ref{Sec:envMass} presents our results, and Section~\ref{Sec:quenching} provides our quenching scenario.
Finally,  Section~\ref{Sec:Discussion} discusses our results in light of the literature and Section~\ref{Sec:summary} wraps up.

\section{Datasets and methods}
\label{Sec:pres}

\subsection{The Horizon-AGN simulation}
The {\sc Horizon-AGN} simulation\footnote{\url{http://www.horizon-simulation.org/}} \citep{dubois14} is a cosmological hydrodynamical simulation suite, implementing the current state-of-the-art of baryonic physics prescriptions at this resolution, down to a scale of $\sim1$~physical~kpc, including AGN feedback.

\subsubsection{Overview}
{\sc Horizon-AGN} is run with a $\Lambda$CDM cosmology with total matter density $\Omega_{\rm  m}=0.272$, dark energy density $\Omega_\Lambda=0.728$, amplitude of the matter power spectrum $\sigma_8=0.81$, baryon density $\Omega_{\rm  b}=0.045$, Hubble constant $H_0=70.4 \, \rm km\,s^{-1}\,Mpc^{-1}$, and $n_s=0.967$ compatible with the WMAP-7 data~\citep{komatsuetal11}.
The size of the simulation box is $L_{\rm box}=100 \, h^{-1}\rm\,Mpc$ on a side, and the volume contains $1024^3$ dark matter particles, corresponding to a dark matter mass resolution of $M_{\rm  DM, res}=8\times 10^7 \, \rm M_\odot$.
Heating of gas from a uniform UV background takes place after redshift $z_{\rm  reion} = 10$ following~\cite{haardt&madau96}.
Gas can cool down to $10^4\, \rm K$ through H and He collisions with a contribution from metals using rates tabulated by~\cite{sutherland&dopita93}. 

Star formation is set to occur in regions of gas number density above $n_0=0.1\, \rm H\, cm^{-3}$ following the Schmidt relation: $\dot \rho_*= \epsilon_* {\rho_{\rm g} / t_{\rm  ff}}$, where $\dot \rho_*$ is the star formation rate mass density, $\rho_{\rm g}$ the gas mass density, $\epsilon_*=0.02$ the constant star formation efficiency, and $t_{\rm  ff}$ the local free-fall time of the gas.
Feedback from stellar winds, supernovae type Ia and type II are included in the simulation, accounting for mass, energy and metal release.
{\sc Horizon-AGN} also follows the formation of black holes, which can grow by gas accretion at a Bondi-capped-at-Eddington rate and coalesce when they form a tight enough binary. 
Black holes release energy in a quasar/radio (heating/jet) mode when the accretion rate is respectively above and below one per cent of Eddington, with efficiencies tuned to match the black hole--galaxy scaling relations at $z=0$~\citep[see][for details]{duboisetal12agnmodel}. 

{\sc Horizon-AGN} has been found to be in fairly good agreement with observations up to $z=4$ \citep[see e.g.,][]{Kaviraj16}.
Some points of tensions remain, especially at low mass, where galaxy masses are overestimated. 
In addition, the separation between star-forming and passive galaxies at low redshift is less pronounced in the simulation than in observations, due to residual star formation even in quenched galaxies.
At high redshift, the high-mass end of the mass function is also slightly underestimated. 
We do not expect that these discrepancies will strongly affect the interpretation of our analysis.

\subsubsection{The galaxy and halo catalogues}\label{Subsec:cat}
\paragraph*{Galaxies} Galaxies are identified using the {\sc AdaptaHOP} structure finder \citep{aubert04,tweedetal09} applied to the distribution of star particles. 
Consistent with previous works of the simulation, structures are selected using a local threshold of 178 times the average matter density, with the local density of individual particles calculated using the 20 nearest neighbours.
Only structures that have more than 50 particles are considered.
Galaxy masses ($M_{\rm st}$) are defined as the sum of all stellar particles in the galaxies, and star formation rates (SFR) are computed by summing the stellar particle masses formed over the last 100 Myrs.

\paragraph*{Haloes} Similarly to galaxies, haloes are identified using the {\sc AdaptaHOP} algorithm. 
The centre of the halo is temporarily defined as the densest particle in the halo, where the density is computed from the 20 nearest neighbours.
In a subsequent step, a sphere of the size of the virial radius is drawn around it and a shrinking sphere method \citep{power03} is used to recursively find the centre of the halo.
In each iteration, the radius of the halo is reduced by 10 per cent.
The search is stopped when a sphere three times larger than our spatial resolution is reached, and the final centre is the densest particle in that sphere.
In the final catalogue, we only consider structures with more than 100 particles and with more than 80 times the average density of the box.
The mass of the halo ($M_{\rm h}$) is defined as the sum of the masses of the member dark matter particles.
We tested that our conclusions are unchanged when the virial mass is adopted instead.
Notice that {\sc AdaptaHOP} identifies a hierarchy of haloes and subhaloes by using information on the local density and the connectivity between the particles.
Subhaloes satisfying the criteria set above are also included in our sample.
Galaxies are then positionally matched to their dark matter haloes.
Those matched to haloes of level$=1$ are central galaxies, and only these central galaxies are kept in our final sample. 
We carefully checked ``splasback galaxies", namely those galaxies which passed by another more massive galaxy but are not bound yet to this new host. They do not dominate the trends that we are measuring. To do so, we performed the same measurements as presented below, but removing splasback galaxies from the sample based on the criterion proposed in \cite{tinkeretal2017}: galaxies with projected separation $R < 2.5\times R_{\rm vir}$ of a larger group and a velocity dispersion $\Delta_{\rm v}<1000 $~km/s with respect to the  central galaxy of the larger group.
This criterion removes only 0.4\% of those classified as centrals by AdaptaHop, and the results with and without splashback galaxies are almost identical. Therefore, we do not consider splashback galaxies separately and remove satellites only for our analysis for simplicity.

\paragraph*{Gas properties within the halo}
In order to interpret the measured environmental trend, we also extract gas properties within the virial radius of the dark matter haloes.
For simplicity, gas leaf cells are turned into particles. 
Gas particle velocities are corrected for the halo peculiar velocity.
In order to investigate gas and angular momentum transfer from the outer part to the inner part of the haloes (i.e., corresponding to the scale of their hosted galaxies), the density and the dimensionless spin parameter of the gas component is then computed in spherical shells \citep[as done e.g. in][]{danovichetal15} of thickness 0.1$\times R_{\rm h,vir}$ around the halo center following the definition \citep{bullock01}:
\begin{equation}
    \lambda=\dfrac{j}{\sqrt{2}R_{\rm h,vir}V_{\rm h,vir}}\, ,
    \label{eqn:spinparam}
\end{equation}
where $j$ is the magnitude of the specific angular momentum (the angular momentum per unit mass) and $R_{\rm h,vir}$ and  $V_{\rm h,vir}$ are the virial radius and the circular velocity at the virial radius respectively \footnote{Note that our measurements of gas density and angular momentum include all gas cells within the halo. For sanity checks, we also tried to identify only gas cells that are gravitationally bound to the halo. To do so, we compute the escape velocity at each radius, knowing the mass of DM, gas and stars enclosed within this radius and assuming spherical symmetry to speed up the computation. Then we keep for our measurements only gas cells with a velocity relative to the galaxy smaller than the escape velocity. This is an approximation, since the matter distribution within the halp is not perfectly spherical and not only gravitational forces are acting on the gas cells. However, when keeping only bound gas cells defined with this method, our conclusions are unchanged.}.

In {\sc Horizon-AGN}, the smallest gas cell size is 1 physical kpc, and this refinement level is reached only in the densest regions within the galaxies. 
In the circumgalactic medium (CGM) however, the cell size  is twice or fourth time larger. At this coarse resolution, gas disks and clumps are barely resolved. 
In addition, gas temperature is not allowed to cool down below $10^4$K to avoid over-cooling issues. 
For these reasons, we limit ourselves  to the measurement of gas density and angular momentum in thick shells\footnote{Each of these shells contains at least 30 leaf cells and on average several hundreds, depending on the halo size and the shell radius.}, without attempting to measure gas temperature or gas disk thicknesses, clumpiness and surface densities, which are all very relevant quantities in general but not meaningful in {\sc Horizon-AGN}. 
We note that because gas angular momentum is primarily built up by torques from the large-scale structure, it should not be strongly affected by this resolution limit.
However, as discussed in \cite{chabanieretal20}, gas angular momentum in {\sc Horizon-AGN} might be slightly overestimated because instabilities are not correctly resolved.
This global overestimation is apparent when comparing the gas spin parameter radial profile in {\sc Horizon-AGN} and results from a higher resolution simulation \citep{danovichetal15}, as shown in Fig.~\ref{fig:profile2}.

\begin{figure}
   \centering
   \includegraphics[width=\columnwidth]{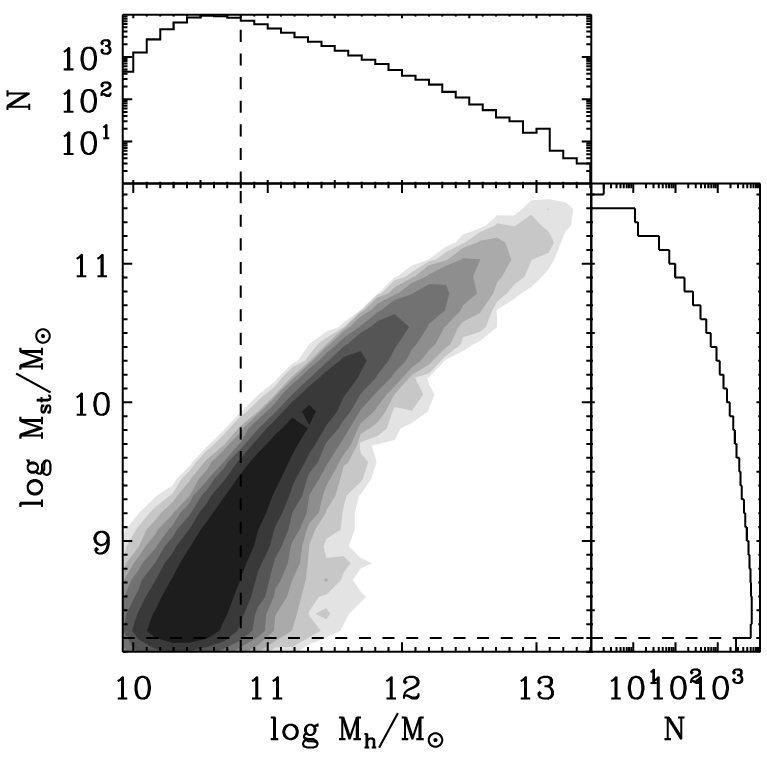}
        \caption{Halo mass $M_{\rm h}$ versus stellar mass $M_{\rm st}$ distribution of central galaxies at $z=1.97$ in {\sc Horizon-AGN}. 
        The vertical and horizontal dashed lines represent the mass cuts adopted in our study.}
\label{fig:AGN_z2_MF}
\end{figure}

\paragraph*{Final sample selection}
We consider in this study only central galaxies, because satellite galaxies are in general known to be affected primarily by their central galaxies \citep[e.g.,][]{weinmannetal2006}.
We also limit our analysis to redshift around two, the epoch of the active star formation and the quenching in action, which helps to highlight the impacts of environment on mass assembly,  whereas at redshift zero many galaxies already become quiescent.
Fig.~\ref{fig:AGN_z2_MF} presents the dark matter halo mass $M_{\rm h}$ versus stellar mass $M_{\rm st}$ distribution for central galaxies at $z=1.97$ in {\sc Horizon-AGN}, along with the mass cuts adopted around the resolution limit of the simulation.
Ultimately, we want to work on a sample limited in terms of halo mass.
We first choose the galaxy mass cut at $10^{8.2}{\rm M}_{\odot}$ as the mass where the galaxy mass function reaches its maximum (corresponding to the completeness limit in the simulation).
We then find a value for the halo mass cut which satisfies completeness of the halo sample given the galaxy mass cut. 
We therefore define the halo mass cut at $10^{10.8}{\rm M}_{\odot}$.
These mass cuts leave 34,356 galaxies. 

\begin{figure*}
   \centering
    \includegraphics[width=0.95\textwidth]{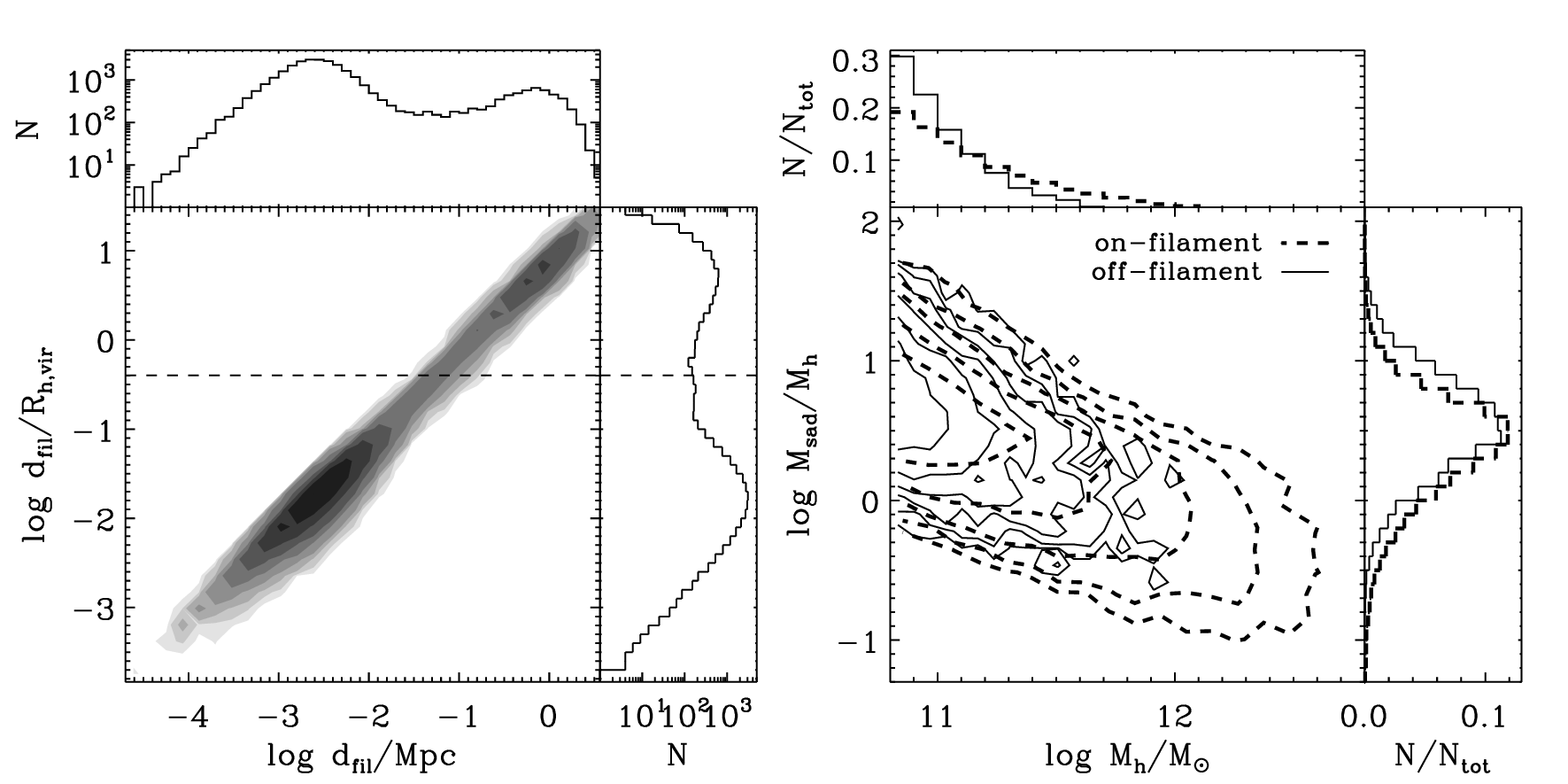}
        \caption{\textit{Left}: $d_{\textrm{fil}}$ versus $d_{\textrm{fil}}/R_{\textrm{h,vir}}$ distribution of central galaxies at $z=1.97$ in {\sc Horizon-AGN}.
        The horizontal \textit{dashed} line shows the cut used to separate ``off-" and ``on-filament" galaxies (i.e., $d_{\rm fil}=0.4 R_{\rm h,vir}\equiv d_{\rm cut}$). 
        \textit{Right}: $M_{\textrm{sad}}/M_{\textrm{h}}$ versus $M_{\textrm{h}}$ contours of ``off-" (\textit{solid} line) and ``on-filament" (\textit{thick dashed} line) haloes.}
\label{fig:AGN_z2_Mh_Msad}
\end{figure*}

\subsection{Defining galaxy environment}
Environment impacts galaxies at different scales.
A comprehensive accounting for the different processes at play requires precisely characterizing the environment at the relevant scales.
As a matter of fact, beyond their dark matter halo, galaxy growth will depend on the proximity, content and kinematics of the large-scale gas reservoir, i.e., the close-by cosmic filaments.
The effect of these filaments might be significant depending on their mass content with respect to the background. 
Beyond associating galaxies to their host halo, we will also compute the local density and distances to filaments and nodes of the cosmic web.

\subsubsection{Local density estimation}
The local density ($\rho$) is estimated from the dark matter particle density field. The Delaunay tessellation of dark matter particles is first computed and then interpolated on a regular grid of resolution 280~kpc, by assigning to each cell of the regular grid the average of the values of the tetrahedra (or their volume fraction) co-existing in this cell. The grid is subsequently smoothed at the scale of~1~Mpc.
The density associated to each galaxy is the density of the grid voxel in which the galaxy is embedded.
In order to confirm that our results are not strongly dependent on the chosen density definition, we also build another density estimator, i.e., the distance to the nearest neighbour galaxy (regardless of if it is a central or a satellite) in order to check that our results about the trend towards cosmic filaments are preserved when changing the density estimator (see Appendix~\ref{Sec:alter}).

\subsubsection{Vorticity estimation}
The vorticity of the gas component is computed for the purpose of illustration in Fig.~\ref{fig:skel} and Fig.~\ref{fig:skel2}. In {\sc Horizon-AGN}, the evolution of the gas field is followed on the AMR grid, and therefore the velocity field is continuous everywhere \citep[unlike for discrete particle distribution, for which measuring accurately the vorticity requires advanced interpolation technique, see e.g.,][]{hahnetal2015}. The AMR grid is projected onto a regular grid, for which the gas velocity vector is known in each cell. The vorticity is then computed as being the curl of the velocity field using fast fourier transform. The bottom panels of Fig.~\ref{fig:skel} and Fig.~\ref{fig:skel2} represent the norm of the gas vorticity vector. Vorticity is confined in massive filaments, and is on overall higher at the edge of the filaments than at the center (two \textit{bottm} panels on the \textit{left} of Fig.~\ref{fig:skel}). This lines up with results from the literature showing that the vorticity within the multi-flow region is mostly distributed near the caustic \citep[e.g.][Ramsoy et al. 2020 submitted]{pichon99, hahnetal2015,laigle2015}.

\subsubsection{Position in the cosmic web}\label{Subsec:CW}
\paragraph*{Filament extraction and classification} Filaments are extracted thanks to the persistence based filament tracing algorithm {\sc DisPerSE} \citep[][]{sousbie111,sousbie112}.
This method identifies ridges from the density field as the special lines connecting topologically robust saddle points to peaks (the so-called ``nodes'').
The identified filamentary network is by construction multi-scale (in that it is defined without having to set a smoothing scale) and the extraction is robust to noise.

The set of all segments defining these ridges is called the skeleton \citep{pogo09}.
In order to determine filament position with the best possible accuracy, we choose to extract our fiducial skeleton {\sc skl}$_{\rm fid}$ directly from the Delaunay tessellation of the \textit{dark matter particle} distribution\footnote{Note that to speed up the computation, the box is divided into 8 sub-boxes and both the Delaunay tessellation computation and the filament extraction are performed on each sub-box individually. Galaxies close to the edges of the sub-boxes are then removed from the subsequent analysis.}.
Note that this choice differs from previous studies investigating the impact of cosmic web while extracting the filaments from the galaxies distribution \citep[e.g.,][]{laigle18}, which suffers from bias and low sampling, or from the gas distribution on a regular grid \citep[e.g.,][]{kraljic18}, which limits the filament position accuracy to the grid resolution.
With the skeleton extracted from the dark matter particle distribution, we can probe very small distance to filaments, at the resolution limit of the simulation. 

Persistence is defined as the ratio of the density value at  two critical points which are topologically paired (e.g., maximum--saddle). Persistence does not define a scale, it defines a significance of the feature over the noise.
A larger persistence threshold isolate the most robust topological features\footnote{Expressed in terms of numbers of $\sigma$  \citep[with an analogy to the gaussian case, see Eq.~(4) in][]{sousbie11}, persistence increases when the probability to find the critical pair  in the Delaunay tessellation of a random discrete Poisson distribution decreases.}.
Since denser filaments are generally defined through more particles, they are  more significant with respect to the particle noise and therefore have a larger persistence. In this sense, a high persistence threshold will tend to privilege the densest, or more ``well-defined" filaments. 

Setting the persistence threshold is left to the user of {\sc DisPerSE}, and the persistence's choice should depend eventually on the characteristics and sampling of the tracers used to estimate density.
In this work, based on dark matter particles to estimate the density, we set persistence  to the conservative threshold of $7\sigma$\footnote{For comparison, \cite{katz2020} and Ramsoy et al. 2020 (submitted), who both identified filaments from the dark matter particle density field, used a persistence threshold of 5$\sigma$ and 10$\sigma$, respectively.}.
On the one hand, we found indeed that a skeleton extraction with a higher threshold would systematically miss some cosmic filaments, which, in turn, would lead to a dilution of the signal when measuring galaxy properties as a function of distance to filaments.
On the other hand, a consequence of further decreasing the persistence threshold below $7\sigma$ would be the inclusion in our set of filaments of more and more small-scale structures (``tendrils"), resulting in mixing the scales for the analysis, which would in turn make the interpretation of the results less straightforward.

As a matter of fact, due to the intrinsically multi-scale nature of the cosmic web network, filaments are diverse in terms of strength and scale, and can impact haloes differently depending on their relative width and density with respect to the size and mass of the haloes. 
The saddle point of each filament provides a natural location in which to estimate its density with respect to the background (being the local density minimum along the filament).
In an attempt to quantify this, we therefore measure the dark matter mass $M_{\rm sad}$ enclosed in a sphere of diameter 1~Mpc around the saddle of each filament. 
Reasonably varying the radius of the sphere around 1~Mpc does not significantly change the result. 
Finally, we use $M_{\rm sad}$ to estimate the strength of the filament.
In the following analysis, we found however that the impact of filaments on galaxies does not depend on the filament strength (at least when all considered filaments have a persistence above $7\sigma$).

\begin{figure*}
   \centering
   \includegraphics[width=0.95\textwidth]{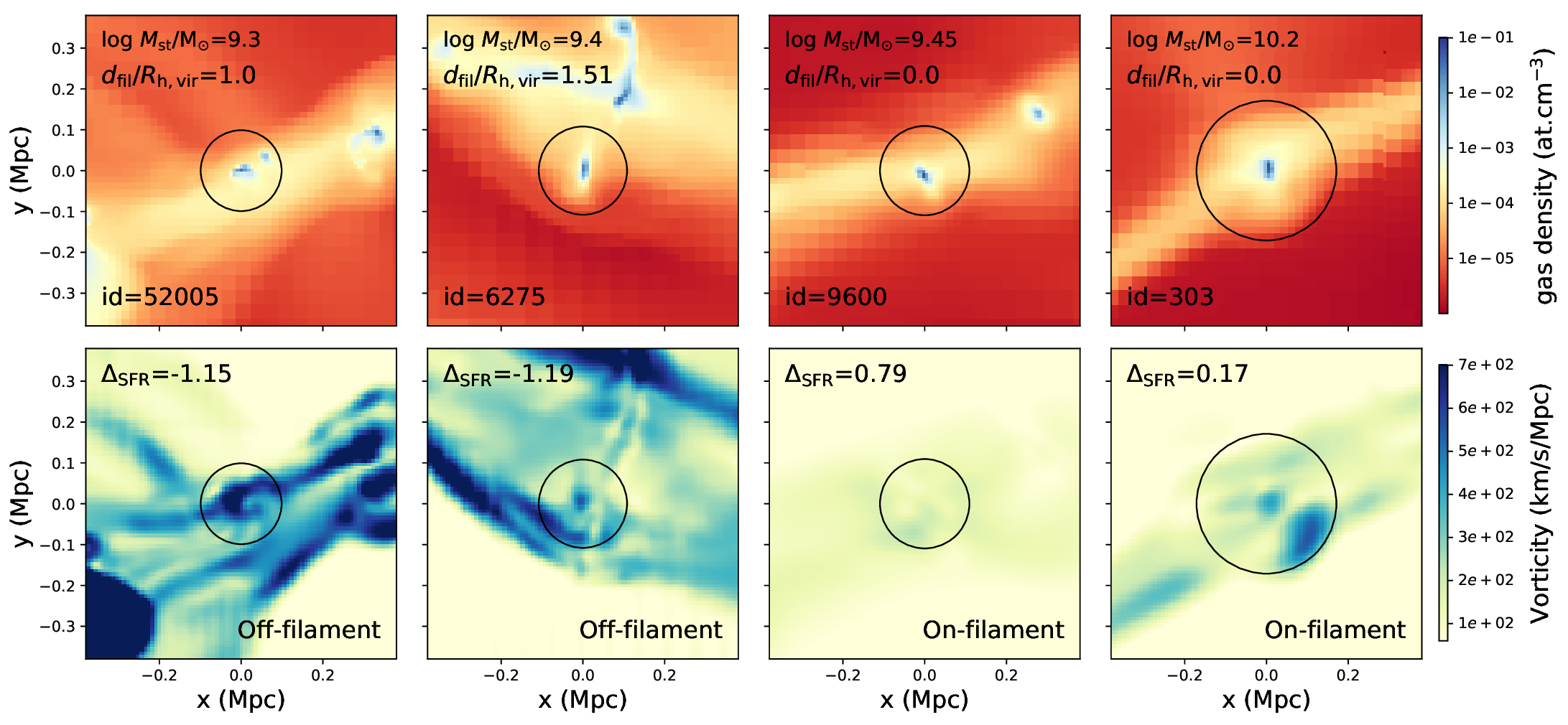}
        \caption{Visualisation of the gas density (\textit{top}) and the norm of gas vorticity (\textit{bottom}) around some of the galaxies in the ``off-" (\textit{two left panels}) and ``on-filament" (\textit{two right panels}) populations in a thin slice.
        The slice is cut in the plane containing the galaxy and its nearest cosmic filament.
        We have chosen the ``off-filament" galaxies sitting \textit{at the vorticity-rich edge} of the filament. 
        Black circles are drawn at the virial radius of each halo.}
\label{fig:skel}
\end{figure*}

\begin{figure}
   \centering
   \includegraphics[width=1.05\columnwidth]{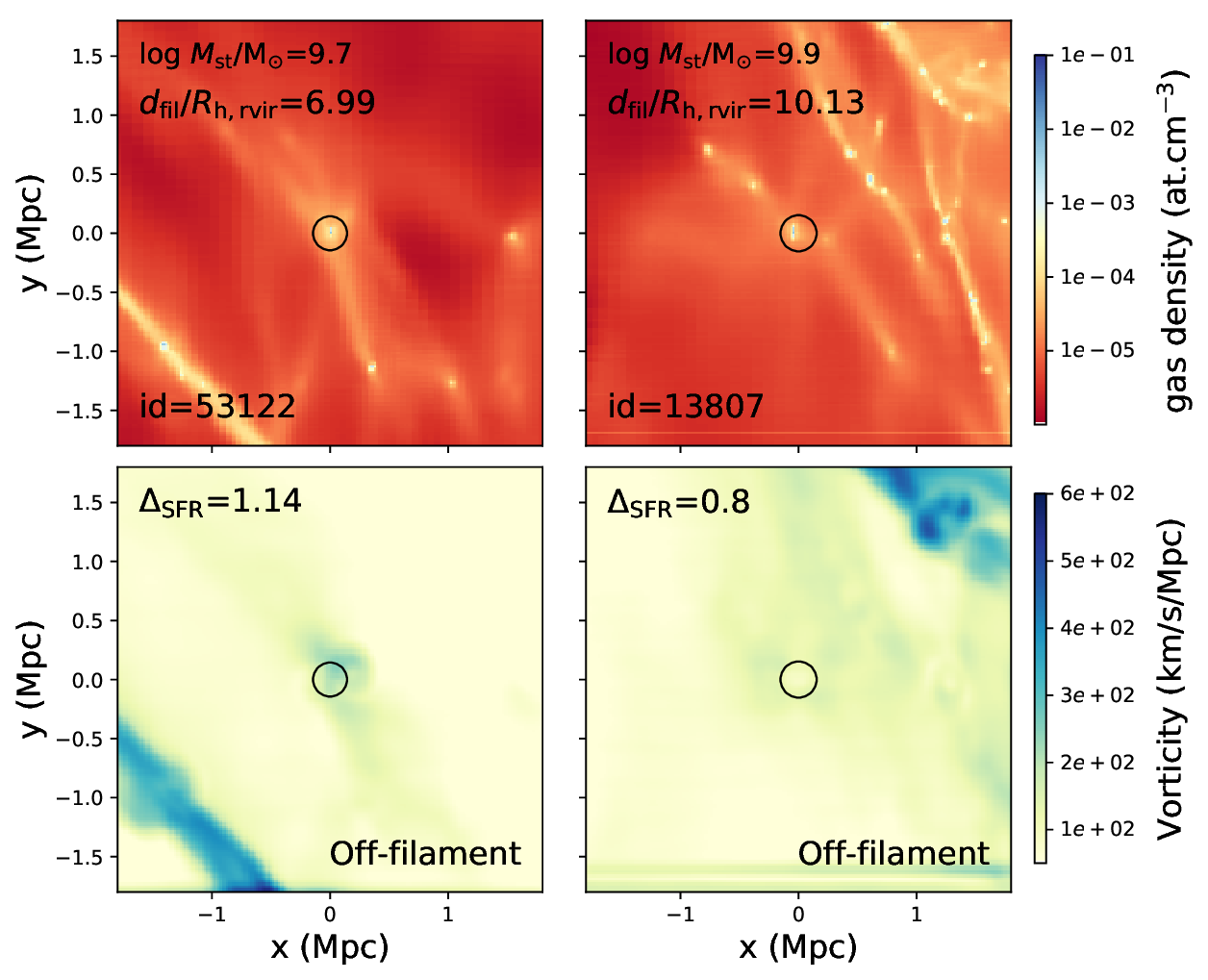}
        \caption{Visualisation of the gas density (\textit{top}) and the norm of gas vorticity (\textit{bottom}) around some of the galaxies in the ``off-filament" sample, in a thin slice.
        The slice is cut in the plane containing the galaxy and its nearest cosmic filament..
        We have chosen these ``off-filament" galaxies sitting \textit{very far} from their closest cosmic filament.
        These galaxies are still connected to gas tendrils with low vorticity content.
        Black circles are drawn at the virial radius of each halo.
        Note that the width of the window (4~Mpc) is larger than on Fig.~\ref{fig:skel} (2~Mpc).}
\label{fig:skel2}
\end{figure}
\paragraph*{Distance to filaments and massive nodes}
As galaxies fall into a cluster, the environment starts to be dominated by the gravitational potential well of the cluster itself; In this case, the specific role played by cosmic filaments is expected to be less important.
Clusters sit at the nodes of the cosmic web estimated from the galaxy or halo distribution \citep[e.g.][]{darraghford19,malavasi2020}. 
As this paper focuses on the role of filaments, we want to exclude all the galaxies in the immediate vicinity of clusters.
Therefore we keep in our sample only galaxies that are further away than 3~Mpc to the  nodes of the cosmic web extracted from the halo distribution, which leaves 31,859 galaxies.

We then compute for each halo the distance $d_{\rm fil}$ to the closest filament in {\sc skl}$_{\rm fid}$.
In the following, galaxy properties will be examined as a function of $d_{\rm fil}/R_{\rm h, vir}$, i.e., the distance to filament \textit{re-normalised} by the virial radius of the halo.
The purpose of this normalization is to account for the fact that dark matter is self-similar and therefore, at first order, the relative distance (depending on its size) of the dark matter halo with respect to the filament is more important than the absolute distance itself. Ideally, we would like 
This normalization has a marginal effect, as can be seen from the distribution of $d_{\rm fil}/R_{\rm h, vir}$ versus $d_{\rm fil}$ in the left panel of Fig.~\ref{fig:AGN_z2_Mh_Msad}.
We checked also that our conclusions do not significantly depend on this normalization.

From Fig.~\ref{fig:AGN_z2_Mh_Msad}, we also see that the distribution of distance to filaments is intrinsically bimodal, with some haloes being ``on-filament", i.e., at very close distance, and others being ``off-filament", i.e., at further distance.
The horizontal \textit{dashed} line on the left panel of Fig.~\ref{fig:AGN_z2_Mh_Msad} shows our threshold to separate the ``on-" and ``off-filament" populations (i.e., $d_{\rm fil}=0.4 R_{\rm h,vir}\equiv d_{\rm cut}$).
This bimodality is physical and arises because, by definition, the spine of the filament will follow the densest ridges of the density field.
Therefore, if a halo does not sit exactly on the filament, the core of the filament has to be denser than the center of this halo (otherwise, the filament would veer off to pass through the halo). 
This defines an ``exclusion zone" radius, which is set up by the rarity of the halo and filament mass \citep{shimetal2020}, and is of the order of the halo virial radius.
Our halo sample being mass-limited, the position of the minimum separating the two modes of the distribution of distance to filaments is broadly set by the minimum halo radius in our sample.
Note that, the dark matter halos being virialized, the exact position of the dark matter filaments passing through the halos is inevitably approximative, as filaments do not formally exist anymore around the halo central regions and are only defined by continuity.
This effect is the reason for the scatter of the distances to the filament for the ``on-filament" halo population (first mode of the distribution displayed in the \textit{left} panel of Fig.~\ref{fig:AGN_z2_Mh_Msad}).

The right panel of Fig.~\ref{fig:AGN_z2_Mh_Msad} shows that the ``on-filament" population (\textit{thick dashed} line) is populated by more massive haloes, and the ``on-filament" population dominates more their filaments (smaller values of $M_{\rm sad}/M_{\rm h}$) than the ``off-filament" population.
In the rest of our work, we consider the mass assembly of these two populations separately.
The ``on-" and ``off-filament" samples consist of 26,495 and 5,364 central galaxies, respectively.
It should be noted that our definitions of ``on-" and ``off-filament" galaxies, based on the position information only, may cause non-negligible mis-classifications. Such confusion could lead to a dilution of the signal when the properties of ``on-" and ``off-filaments" galaxies differ significantly.
The classification would be greatly improved if the velocity information of galaxies relative to filaments is used as well.
However, for simplicity, we will use the classification based on the position information only here.

For illustration, Fig.~\ref{fig:skel} displays the gas density and the norm of the vorticity vector in a thin slice around some of the haloes in our ``on-" (\textit{two right} panels) and ``off-filament" (\textit{two left} panels) samples with similar halo mass and local density.
The slice is cut in the plane containing both the galaxy and its nearest cosmic filament.
The ``off-filament" galaxies are chosen among those in the direct vicinity (i.e., $d_{\rm fil}\sim d_{\rm cut}=0.4 R_{\rm h,vir}$) of a cosmic filament. 
For the purpose of our upcoming arguments in Section \ref{Sec:quenching}, we additionally present some examples of ``off-filament" galaxies  which are \textit{far away} (i.e., $d_{\rm fil} \gtrsim d_{\rm char}=4 R_{\rm h,vir}$, which will be introduced later in Section \ref{Sec:envMassGal}) from their closest cosmic filament in Fig.~\ref{fig:skel2}.
It is important to note that these galaxies, although not identified as sitting on a cosmic filament from our filament extraction, appear to be still connected to gas tendrils. 

\begin{figure*}
   \centering
   \includegraphics[width=0.8\textwidth]{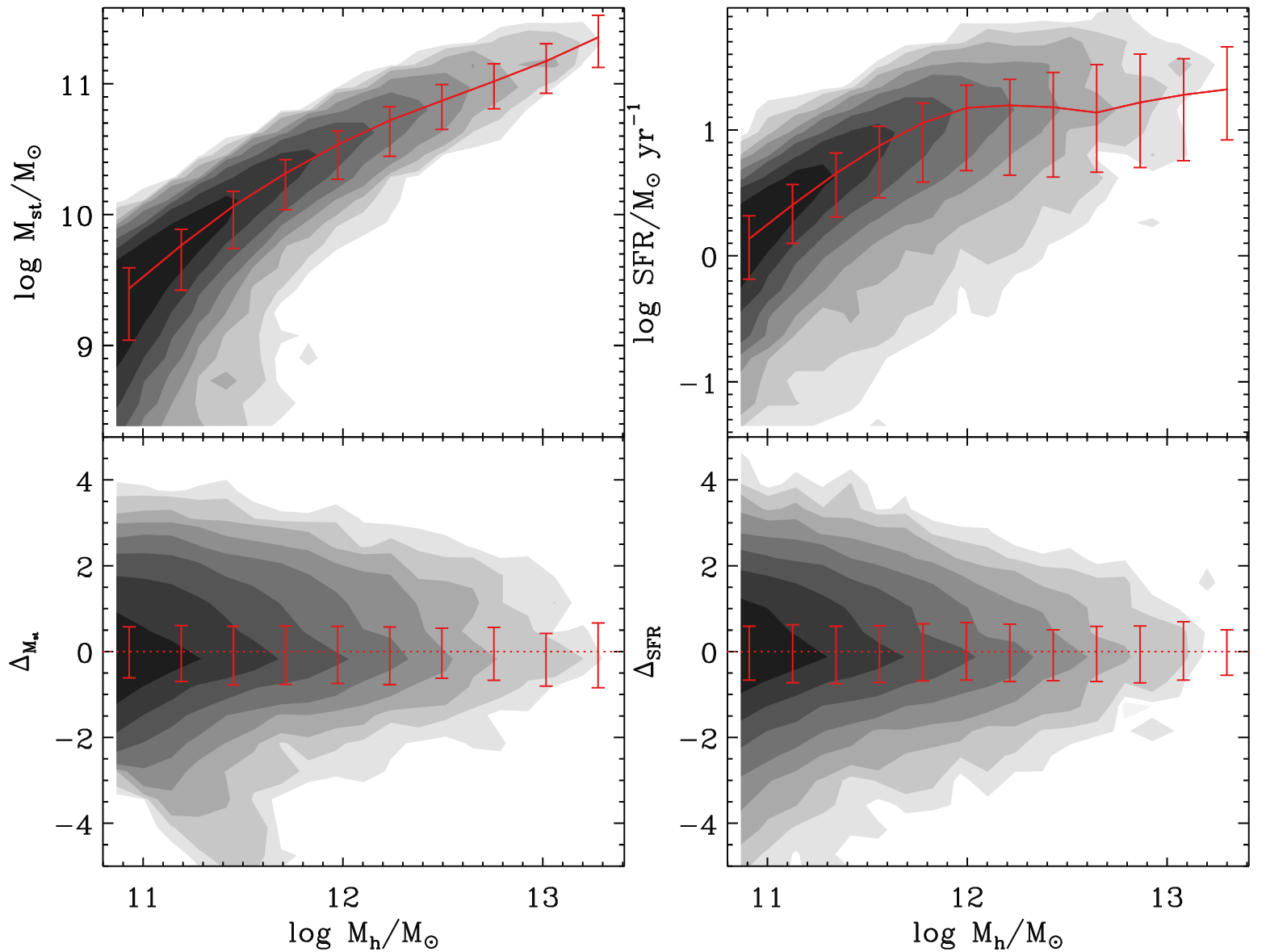}
        \caption{Stellar mass (\textit{left}) and SFR (\textit{right}) and their residuals (\textit{bottom} panels) as a function of halo mass for central galaxies in {\sc Horizon-AGN} at $z=1.97$.
        Residuals are computed as the offset of a given property (mass or SFR) from its typical value at a given $M_{\rm h}$ (red lines in the \textit{top} panels), normalised by the standard deviation $\sigma$.
        The red dotted lines in the \textit{bottom} panels are the zero residual and the error bars are 1$\sigma$ scatter of the residuals measured separately for positive and negative residuals.}
\label{fig:AGN_z2_Mh_propres}
\end{figure*}

\section{Methodology of our measurement}
\label{Sec:methodology}
The heart of our analysis consists in disentangling the potentially distinct effects of halo mass, local density and proximity to filaments in shaping galaxy properties.
To this end, we control first for the effect of halo mass by computing residuals of galaxy properties with respect to the median value at a given halo mass $M_{\rm h}$.
Recall that we want to understand what is the specific role of local density and proximity to filaments in driving the scatter in e.g., the $M_{\rm h}$ versus $M_{\rm st}$ relation.
Galaxy properties and residuals are then analyzed in two-dimensional diagrams, as a function of local density and distance to filaments. 

\begin{figure*}
   \centering
   \includegraphics[width=0.9\textwidth]{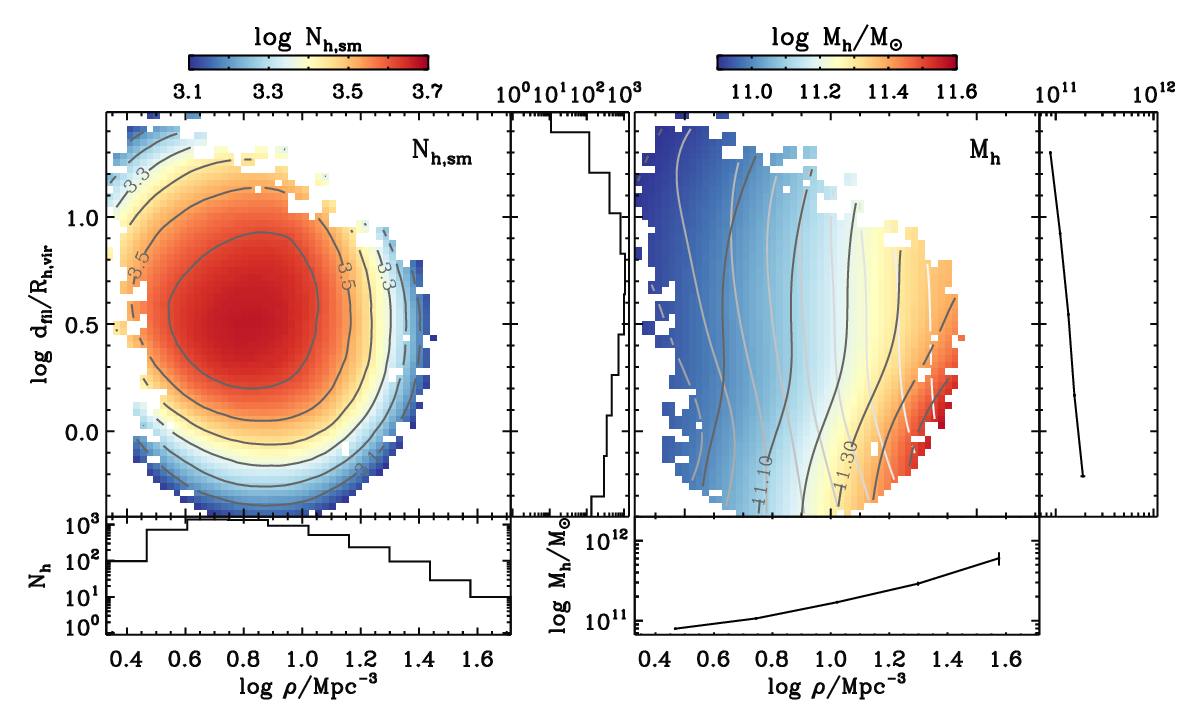}
        \caption{\textit{Left:} Two-dimensional map with dark contours displaying the number of objects used to calculate the weighted mean of a quantity of interest in a given bin of local density ($\rho$)--distance to filaments ($d_{\rm fil}$/$R_{\rm h, vir}$) for haloes that are ``off-filaments" (i.e. $d_{\rm fil}\geq0.4R_{\rm h,vir}$) of {\sc Horizon-AGN} at $z=1.97$.
        Two small panels on the \textit{right} and \textit{bottom} show the histograms of $d_{\rm fil}$/$R_{\rm h,vir}$ and $\rho$, respectively.
        \textit{Right:} Mean halo mass in the local density--distance to filaments plane.
        Halo mass increases both as a function of distance to filaments and density.
        Grayscale contours denote different strengths of filaments, probed by $M_{\rm sad}$, the dark matter mass around the saddle.
        The whiter the color, the stronger the filament.
        In the small panels, the mean halo mass are presented.}
\label{fig:AGN_z2_OffFil_rhodfil_Mh}
\end{figure*}

\subsection{Removing the effect of halo mass}\label{Subsec:residual}
In order to disentangle the effect of environment (local density or proximity to filaments) from the effect of halo mass, we compute the ``residual" of a property for each galaxy with respect to the typical value at its halo mass. 
To do so, we first look at the distribution of the property of interest (e.g., stellar mass) as a function of halo mass.
Among mean, median, and mode values of the property in a given halo mass bin, we choose the most representative of the bulk of the galaxy population. 
For example, when a distribution is not symmetric but skewed, its mode rather than mean or median is more representative.
Because the distribution is in general asymmetric, the mode is chosen to define the main relation between the property of interest and halo mass.
We emphasise that, although we remove galaxies in the vicinity of massive nodes and consider separately the ``on-" and ``off-filament" populations, the main relation is measured based on all galaxies.

Once the main relation is derived, the residual is computed as the deviation from the main relation.
Here, the deviation is measured in terms of $1\sigma$ scatter\footnote{It is computed as the mean of squared deviations from the main relation.} of the property in the corresponding halo mass bin.
The $1\sigma$ is computed separately for the galaxies above and below the main relation (i.e., asymmetric scatters).
The resulting residual values depend slightly on halo mass binning.
We carefully adjust the binning to trace the majority of galaxies as closely and smoothly as possible.

Fig.~\ref{fig:AGN_z2_Mh_propres} presents $M_{\rm st}$ and SFR (\textit{top} panels) of central galaxies in {\sc Horizon-AGN} at $z=1.97$ as a function of $M_{\rm h}$ along with their residuals (\textit{bottom} panels).
$M_{\rm st}$ and SFR show a clear dependence on $M_{\rm h}$.
However, the distributions of their residuals are peaked at the zero residual with similar scatters over all $M_{\rm h}$ bins with no $M_{\rm h}$ dependency, suggesting that the $M_{\rm h}$ effect is well removed.
Therefore, with such residual quantities, we can investigate environmental effects on galaxy properties beyond halo mass.
In addition, we can safely stack galaxies of different halo masses, which is critical to increase the statistical significance of any environmental effects that are in general expected to be subtle and therefore more difficult to highlight with low statistics.
It should be noted that the asymmetric distribution of a quantity for a fixed halo mass results in non-zero mean or median of residuals, which is indeed visible in figures shown later.

\subsection{Two-dimensional diagrams}
In order to separate the effect of local density from the effect of proximity to filaments, we examine galaxy properties in a two-dimensional plane defined by local density and distance to filaments.
At a given position in the two-dimensional plane, we calculate a spline kernel weighted mean for the property of interest.
The spline kernel we adopt is from \citet{monaghan&lattanzio85}:
\begin{equation}
W(r_i,h_{\textrm{spl}})=\frac{1}{\pi h_{\textrm{spl}}^3}
\begin{cases}
1-\frac{3}{2}q_i^2 + \frac{3}{4}q_i^3 \,; q_i\le1 \\
\frac{1}{4}(2-q_i)^3 \,; \qquad 1\le q_i\le2 \\
0 \,; \qquad\qquad\qquad \text{otherwise},
\end{cases}
\label{eqn:spline}
\end{equation}
where $r_i$ is the distance between the position where the spline kernel weighted mean is calculated and the position of galaxy $i$ in the plane, $q_i=r_i/h_{\textrm{spl}}$, and $h_{\textrm{spl}}$ the smoothing length.
We divide the two-dimensional plane into 60$\times$60 bins, and the weighted mean is calculated in each bin with a smoothing length ($h_{\textrm{spl}}$) that corresponds to 11 bins along an axis.
We keep bins for which at least 1000 galaxies are used to calculate the weighted mean (i.e., galaxies with non-zero spline kernel weights).

\begin{figure*}
   \centering
   \includegraphics[width=0.7\textwidth]{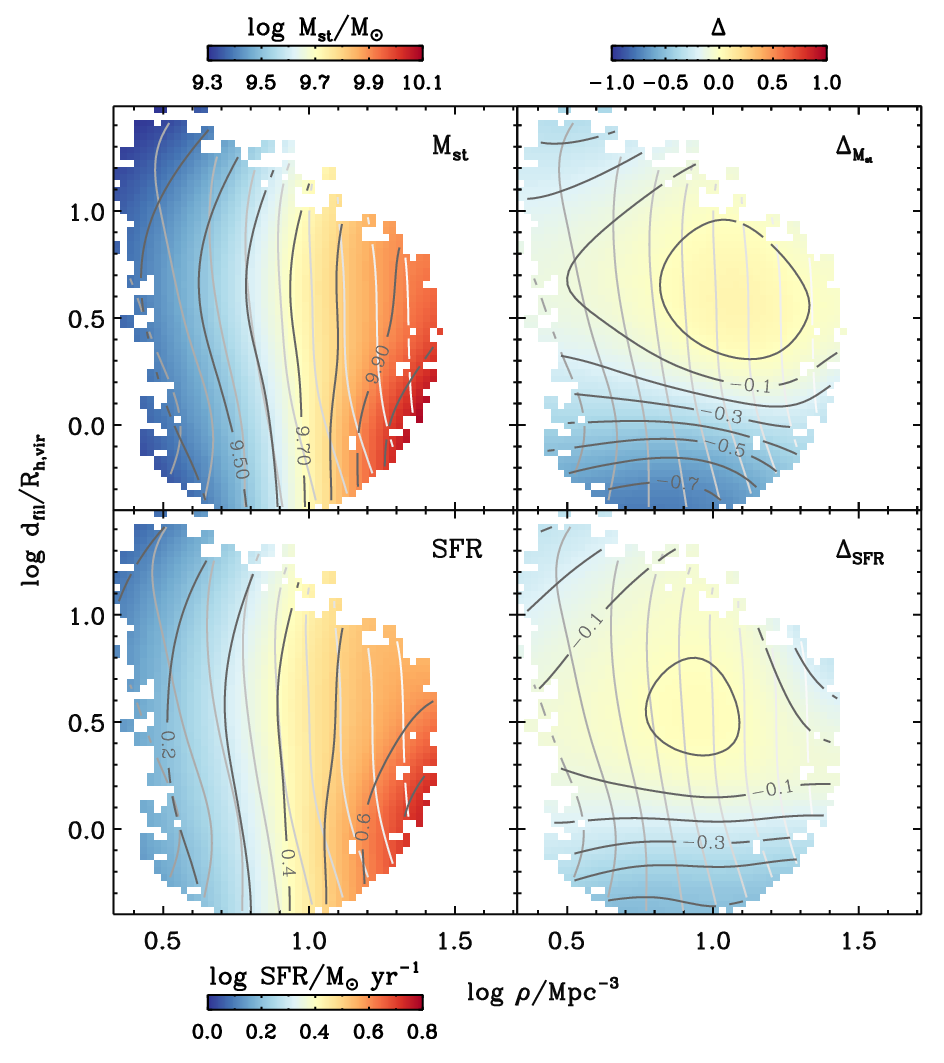}
        \caption{\textit{Top:} Stellar mass (\textit{left}) and its residual (\textit{right}) in the plane of distance to filaments ($d_{\textrm{fil}}$/$R_{\textrm{h,vir}}$) versus local density ($\rho$) for haloes that are ``off-filaments" for Horizon-AGN at $z=1.97$. 
        \textit{Bottom:} SFR (\textit{left}) and its residual (\textit{right}) in the same plane. 
        After removing the effect of halo mass (residuals), stellar mass and SFR gradients are mainly driven by distance to filaments and not by density, being positive and then negative as approaching filaments.
        Grayscale contours denote different strengths of filaments, probed by $M_{\rm sad}$, the dark matter mass around the saddle.}
\label{fig:AGN_z2_OffFil_rhodfil_Mst_SFR}
\end{figure*}

\section{How environment shapes masses and SFR}
\label{Sec:envMass}
In this section, we investigate halo mass, galaxy mass, SFR, and their residuals as a function of local density and distance to filaments. 
We first display properties of the ``off-filament" population in the plane of local density ($\rho$) and distance to filaments (normalised by halo virial radius; $d_{\rm fil}/R_{\rm h,vir}$) in Sections \ref{Sec:envMassHalo} and \ref{Sec:envMassGal}.
Then, we compare ``off-" and ``on-filament" populations in Section \ref{Sec:SFOnVSOff}.

\subsection{Halo mass assembly as a function of environment for ``off-filament haloes"}\label{Sec:envMassHalo}
On the \textit{left} panel of Fig.~\ref{fig:AGN_z2_OffFil_rhodfil_Mh}, we present the number of ``off-filament" haloes used to calculate the weighted mean of the quantity of interest for a given bin in the local density--distance to filaments plane. 
The \textit{right} panel shows the same plane color-coded by the weighted mean of $M_{\rm h}$ (see equation \eqref{eqn:spline}). 
\textit{Black} line in the \textit{right} and \textit{bottom} sub-panels represents the one-dimensional mean relation as a function of distance to filaments and local density, respectively.
Halo mass gradient is clearly a function of both local density and distance to filaments; the $M_{\rm h}$ contours (\textit{black} contours) are not parallel to any of the $x$- and $y$-axes, and the median or mean halo mass increases both with local density and distance to filament. 
This finding is in agreement with previous works from the literature. 
\cite{Mussoetal2017} demonstrated that assembly bias is influenced by the tides of the cosmic web.
At fixed density, the presence of a nearby anisotropic proto-filament further shapes the halo mass function beyond local density \citep[see also][]{kraljic19a}.

To get a more precise idea of how halo mass gradient varies when approaching \textit{a given filament}, we present \textit{grayscale} lines in the \textit{right} panel of Fig.~\ref{fig:AGN_z2_OffFil_rhodfil_Mh} of constant $M_{\rm sad}$ (the whiter the color, the denser the filament). 
The slope of these lines in the $d_{\rm fil}/R_{\rm h,vir}$ versus $\rho$ plane depends on the smoothing scale used to estimate the density $\rho$ and the radius used to measure the mass at the saddle $M_{\rm sad}$.
Formally speaking, mass gradients towards filaments, as discussed in numerous observational works, arise along these lines, and not at fixed density along the vertical axis.
When approaching a filament (following a given grayscale line), halo mass increases as expected.
The strength of the gradient (how many mass contours are crossed) depends on the strength of the filament; a larger gradient is driven by a stronger filament.

The dependency of halo mass on both local density and distance to filaments supports the idea of adopting the residual quantities (see Section~\ref{Subsec:residual}) to explore the direct impact of environment on galaxies \textit{beyond halo mass}.
In the following sections, we will present the residuals of galaxy properties along with the properties themselves.

\subsection{Galaxy mass assembly as a function of environment for ``off-filament galaxies"}\label{Sec:envMassGal}
We now investigate galaxy mass assembly in this local density--distance to filaments plane. 
The \textit{top left} panel of Fig.~\ref{fig:AGN_z2_OffFil_rhodfil_Mst_SFR} presents $M_{\rm st}$ gradients.
These contours are relatively similar to the $M_{\rm h}$ contours displayed in the right panel of Fig.~\ref{fig:AGN_z2_OffFil_rhodfil_Mh}, indicating that most of the dependency of $M_{\rm st}$ on environment occurs through the $M_{\rm st}$--$M_{\rm h}$ relation and the dependency of $M_{\rm h}$ on environment. 

Since we are interested in exploring if environment plays a role beyond what is driven by $M_{\rm h}$, the \textit{top right} panel of Fig.~\ref{fig:AGN_z2_OffFil_rhodfil_Mst_SFR} presents the weighted mean of stellar mass residual ($\Delta_{M_{\rm st}}$). 
$\Delta_{M_{\rm st}}$ changes within $-0.8\sigma$ and 0.1$\sigma$.
That is, about 34\% ($1\sigma$) of the diversity in stellar mass for a given halo mass is explained by local density and distance to filaments.
On the other hand, this also suggest that these two environmental parameters considered here are not the only driver of the scatter in the $M_{\rm st}$ versus $M_{\rm h}$ relation.
However, there are some caveats.
The signal might be naturally underestimated because of the smoothing through the computation of the spline kernel weighted mean performed on this plane.
Inaccuracies in the extraction of the skeleton can further blur the trend. 
More statistics on possibly higher resolution simulations would be needed to improve the signal.
The reason why the weighted mean residual tends to be negative is because the distribution of $M_{\rm st}$ in a given $M_{\rm h}$ bin is not symmetric (see Fig.~\ref{fig:AGN_z2_Mh_propres} and Section \ref{Subsec:residual}).

In the stellar mass residual map, we clearly measure an enhancement of $\Delta_{M_{\rm st}}$ when approaching filaments, the peak value being reached at $\log d_{\rm fil}/R_{\rm h,vir}\sim0.6$ or $d_{\rm fil}\sim 4\times R_{\rm h,vir}$ (let's call it $d_{\rm char}$). 
At closer distances, $\Delta_{M_{\rm st}}$ decreases as approaching filaments.
This decrement is systematic at all densities (i.e., whatever the filament strength is).
This residual trend can also be described in terms of stellar-to-halo mass ratio. 
At $d_{\rm fil}>d_{\rm char}$, the stellar-to-halo mass ratio increases (increasing $\Delta_{M_{\rm st}}$) as approaching filaments, an indication that the galaxy mass assembly becomes more efficient.
Then, at $d_{\rm fil}<d_{\rm char}$, the trend reverses, an indication that galaxies assemble less efficiently their mass.
We checked that the measured signal is not dependent on halo mass estimators, by re-doing the analysis using either the virial dark matter mass or the peak halo mass\footnote{i.e. the maximum mass reached during the halo life, computed along each halo merger tree.} instead of the total halo mass.
We note that when controlling for the effect of halo mass, most of the dependency of $M_{\rm st}$ \textit{on local density} disappears and the strongest trend is driven by distance to filaments.

We now turn to investigate instantaneous mass assembly, i.e., SFR gradients in this two-dimensional plane.
As displayed in the \textit{bottom} panels of Fig.~\ref{fig:AGN_z2_OffFil_rhodfil_Mst_SFR},
SFR and residual SFR ($\Delta_{\rm SFR}$) present similar behaviour to stellar mass.
This is expected from the correlation between stellar mass and SFR.

\cite{kraljic19a} looked at galaxy mass gradients in the oriented plane of the saddle.
They found that galaxies are more massive closer to the filament spine than further away.
It should be noted however that, for technical reasons, their analysis was carried out at fixed scale.
Their smallest distance bin encompasses both all the data in our plot below $\log d_{\rm fil}/R_{\rm h,vir}\sim 0.3$ \textit{and} the ``on-filament" population, which would veil the reduction of the efficiency of galaxy mass assembly at close distances.
In fact, we caution that such a difficulty is present more widely in every work aiming at probing environmental trends by simply separating galaxies in and out filaments.
The trend being subtle and non-linear, it becomes very easy to mix-up different populations, leading to a non-detection or an opposite signal \citep[e.g.,][]{Martizzietal2020}.

\begin{figure*}
   \centering
   \includegraphics[width=0.7\textwidth]{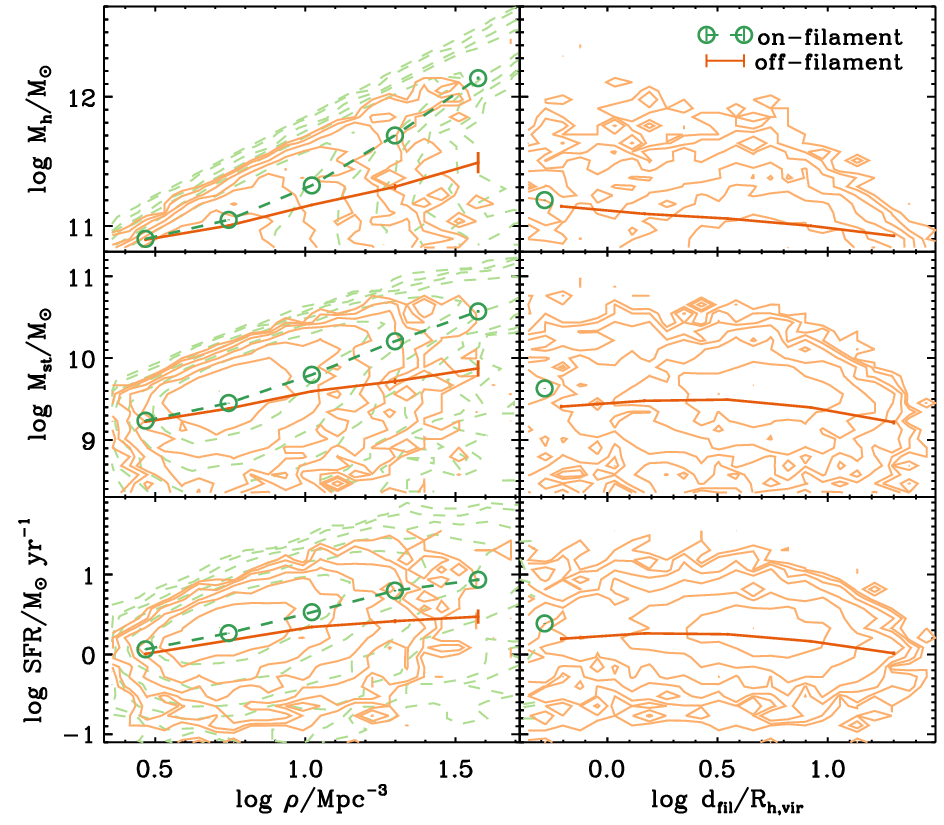}
        \caption{Mean halo mass (\textit{top} panels), stellar mass (\textit{middle}), and SFR (\textit{bottom}) of ``off-" (\textit{orange} \textit{solid lines} with error bars) and ``on-filament" (\textit{green} \textit{open circles with dashed lines}) populations as functions of local density (\textit{left}) and distance to filaments (\textit{right}), respectively.
        The contours with thinner lines and lighter colors show overall distributions of ``off-" (orange solid lines) and ``on-filament" (green dashed lines) populations.
        Although halo mass increases smoothly as the galaxies are closer to filaments until they are ``on-filament" (\textit{top right} panel), note how different are stellar masses and SFR for the ``on-" and ``off-filament" populations (compare green and orange symbols in the \textit{middle and bottom right} panels).
        The errors are computed from bootstrap resamplings.
        The error bars for ``on-filament" are very small.}
\label{fig:AGN_z2_1DMap}
\end{figure*}

\begin{figure*}
   \centering
   \includegraphics[width=0.7\textwidth]{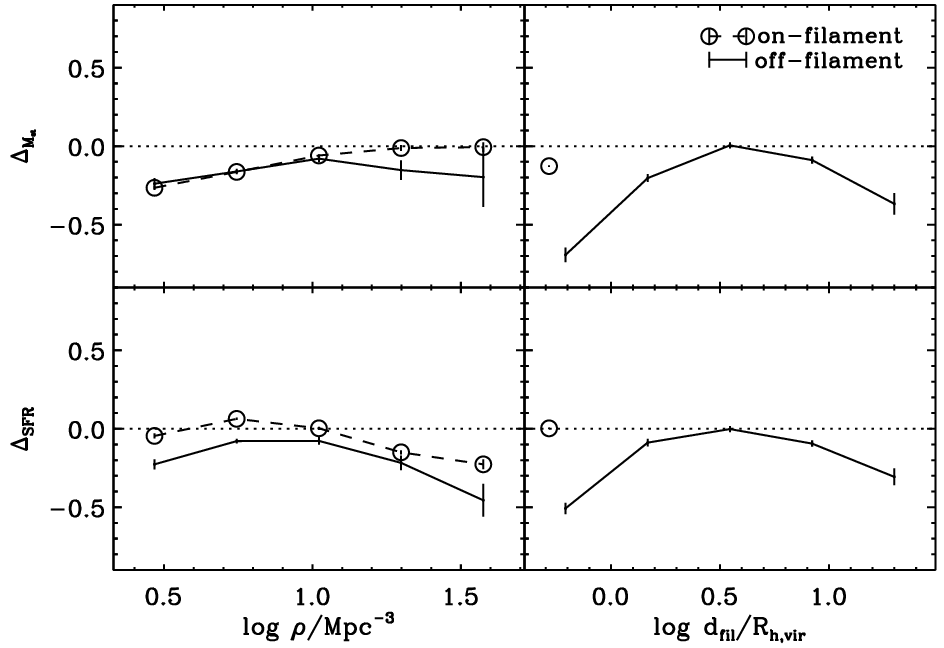}
        \caption{Mean stellar mass residual (\textit{top}) and mean SFR residual (\textit{bottom}) of ``off-" (\textit{solid lines} with error bars) and ``on-filament" (\textit{open cicles with dashed lines}) populations as functions of local density (\textit{left}) and distance to filaments (\textit{right}), respectively. 
        stellar mass and SFR residuals exhibit a strong trend as a function of distance to filaments, peaking at a characteristic distance ($d_{\rm char}$ defined in Section \ref{Sec:envMassGal}) and decreasing in the direct vicinity of the filament spines.
        The errors are computed from bootstrap resamplings.
        The error bars for ``on-filament" are very small.}
\label{fig:AGN_z2_1DMap_res}
\end{figure*}

\subsection{``On"- and ``off-filament" populations}\label{Sec:SFOnVSOff}
So far we have examined the change in galaxy populations as getting closer to filaments.
In this section, we compare the properties and residual properties of the ``off"- and ``on-filament" galaxy populations.
Fig.~\ref{fig:AGN_z2_1DMap} shows halo mass (\textit{top} panel), stellar mass (\textit{middle}), and SFR (\textit{bottom}) of the ``off-" (\textit{solid line} with error bars) and ``on-filament" (\textit{open circles with dashed line}) populations as functions of local density (\textit{left}) and distance to filaments (\textit{right}), respectively.
\textit{Black} and \textit{red} colours correspond to median and mean, respectively.
The errors are computed from bootstrap resamplings.
Halo mass, stellar mass, and SFR increase as local density increases and distance to filaments decreases, as seen in the previous figures (\textit{right} panel of Fig.~\ref{fig:AGN_z2_OffFil_rhodfil_Mh} and \textit{left} panels of Fig.~\ref{fig:AGN_z2_OffFil_rhodfil_Mst_SFR}).

On the \textit{left} panels (i.e., dependence on local density), ``on-filament" halo masses show a stronger slope, being above the trends of ``off-filament" haloes across the whole density range.
This suggests that down to $z=1.97$, halo mass and halo accretion rate are enhanced (at fixed density) when haloes sit exactly on the cosmic filaments, i.e., at the bottom of the gravitational potential well, than further a way, a picture once again consistent with \cite{Mussoetal2017}.
Enhanced stellar masses and SFR for the ``on-filament" populations might simply arise from this halo trend. 

The \textit{right} panels show that the changes of halo mass, stellar mass, and SFR with distance to filaments are rather mild compared to those as a function of local density.
The halo mass of the ``on-filament" population smoothly follows the trend of the ``off-filament" population; these haloes that lie exactly on filaments are those which grow the most.
However, stellar mass and SFR do not show such a smooth transition from the ``off"- to the ``on-filament" populations; the stellar mass and SFR of the ``off-filament" population reach a plateau when approaching filaments, which is followed by a sudden jump for the ``on-filament" population. 
This implies that the mass assembly efficiency of dark and stellar material diverges at the boundaries of filaments.

This discontinuous transition between the ``off-" and the ``on-filament" populations is more dramatic when looking at the residuals of stellar mass and SFR (Fig.~ \ref{fig:AGN_z2_1DMap_res}).
Both the stellar mass and SFR residuals increase as distance to filaments decreases, peak at $d_{\rm char}$, and decrease toward their minima down to the very close distance from filaments (i.e., $d_{\rm cut}$).
Then, when galaxies are exactly at the bottom of the gravitational potential well of filaments (i.e., the ``on-filament" population), their stellar mass and SFR residuals jump again to large values which are comparable to their peak values.
This drastic change, compared to the smooth and continuous halo mass growth as they approach filaments, emphasises different responses of stellar and dark matter as a function of distance to filaments.

From our findings we conclude that there must exist a quenching mechanism, specific to filaments and acting at close distances to the filament spine, which prevents galaxies from forming stars and assembling their mass as efficiently as they in principle should given their host halo mass.
In order to shed light on this mechanism, we will investigate the properties of gas and the morphology of the stellar distribution in the next section.

\section{Understanding galaxy quenching near filaments}
\label{Sec:quenching}
Let us summarize our findings. 
Halo mass increases both as functions of local density and inverse distance to filament, a trend in agreement with theoretical expectations. 
Stellar mass displays a similar dependency overall, as we expect from the relation between stellar and halo mass. 
However, after having controlled for halo mass, we found that galaxy mass assembly, as traced both by stellar mass and SFR, is more efficient down to a characteristic distance to cosmic filaments $d_{\rm char}$, and then proceeds less efficiently when further approaching.
This finding is robust against the definition of halo mass and local density estimators (see Appendix \ref{Sec:alter}), and persists at all filament strengths.
Let us now examine the gas content inside haloes and the morphology of the stellar distribution in order to understand underlying mechanisms that may cause the trend with distance to filaments.  

\begin{figure*}
   \centering
   \includegraphics[width=\textwidth]{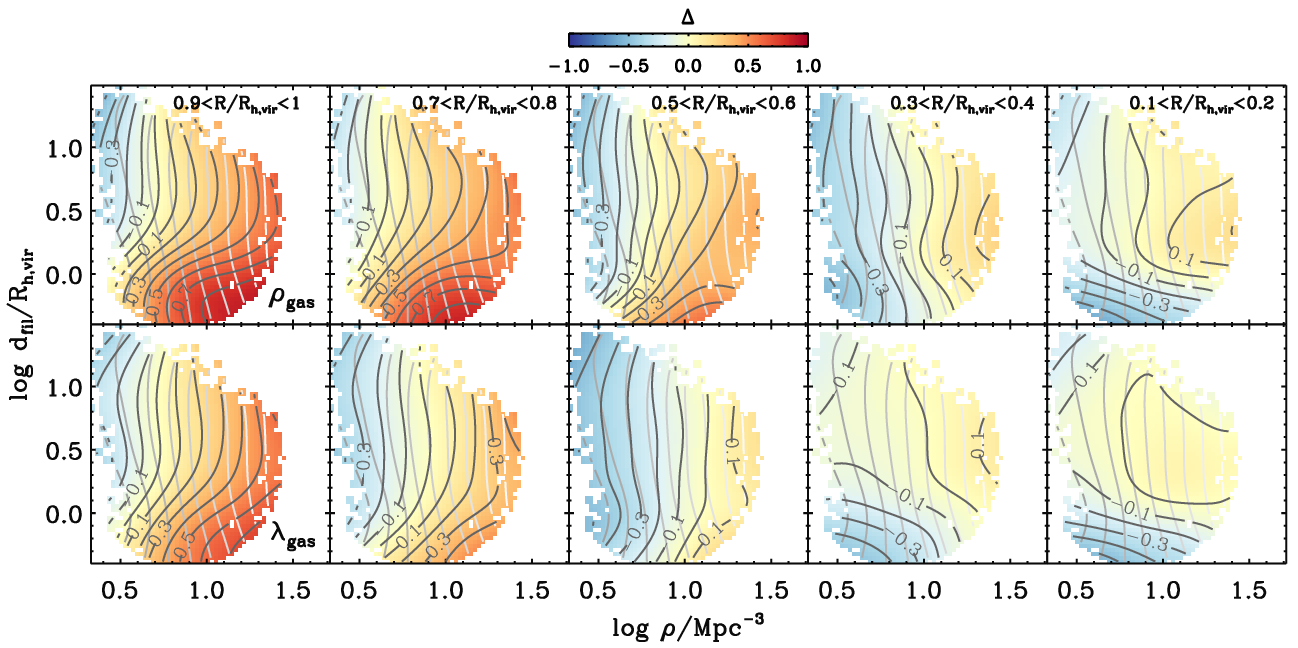}
        \caption{Gas density residuals (\textit{top}) and spin parameter residuals (\textit{bottom}) in shells of thickness 0.1$\times R_{\rm h,vir}$ and radii varying from 0.9$\times R_{\rm h,vir}$ to 0.1$\times  R_{\rm h,vir}$ (from \textit{left} to \textit{right}) in the plane of distance to filaments ($d_{\textrm{fil}}$/$R_{\textrm{h,vir}}$) and local density ($\rho$) for ``off-filament" haloes.
        Grayscale contours denote different strengths of filaments, probed by $M_{\rm sad}$, the dark matter mass around the saddle.}
\label{fig:AGN_z2_OffFil_rhodfil_rhogas_jgas}
\end{figure*}

\begin{figure*}
    \centering
    \includegraphics[width=0.9\textwidth]{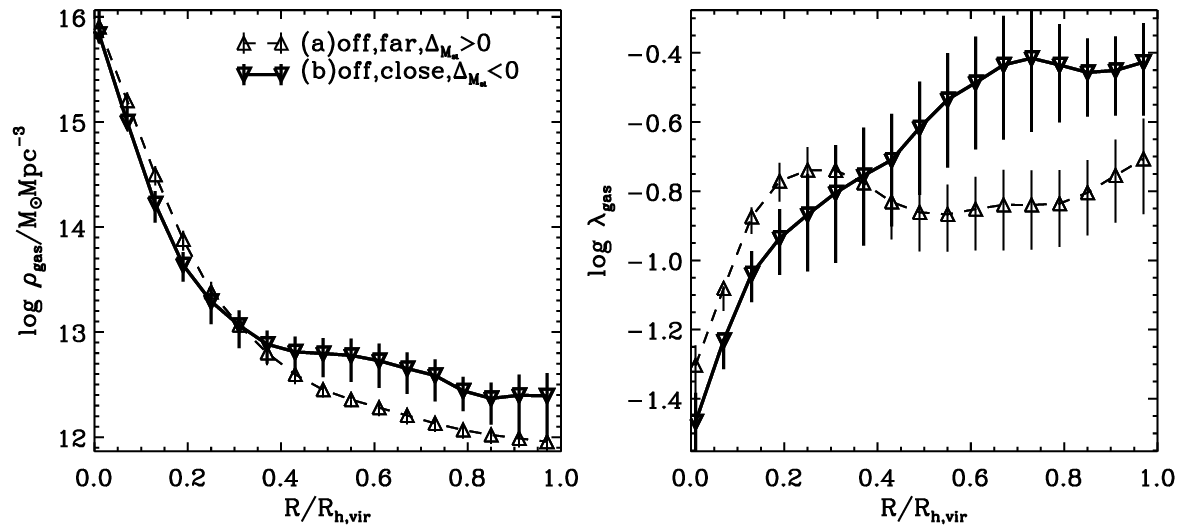}
    \caption{Averaged gas density (\textit{left}) and gas spin parameter (\textit{right}) profiles as a function of distance to the halo center for samples of (a) 27 ``off-filament" galaxies far from filaments (i.e., around $d_{\rm char}$) with positive stellar mass residuals (\textit{dashed lines with triangles}) and of (b) 20 ``off-filament" galaxies close to filaments (i.e., around $d_{\rm cut}$) with negative stellar mass residuals (\textit{thick solid lines with inverted triangles}).
    In both cases, galaxies are taken in a similar bin of halo mass ($11.3< {\rm log} M_h/{\rm M}_{\odot} <11.6$) and local density ($1.0< {\rm log} \rho/{\rm Mpc}^{-3} <1.5$).
    The errors are computed from bootstrap resamplings.}
    \label{fig:profile}
\end{figure*}

\begin{figure*}
   \centering
   \includegraphics[width=0.9\textwidth]{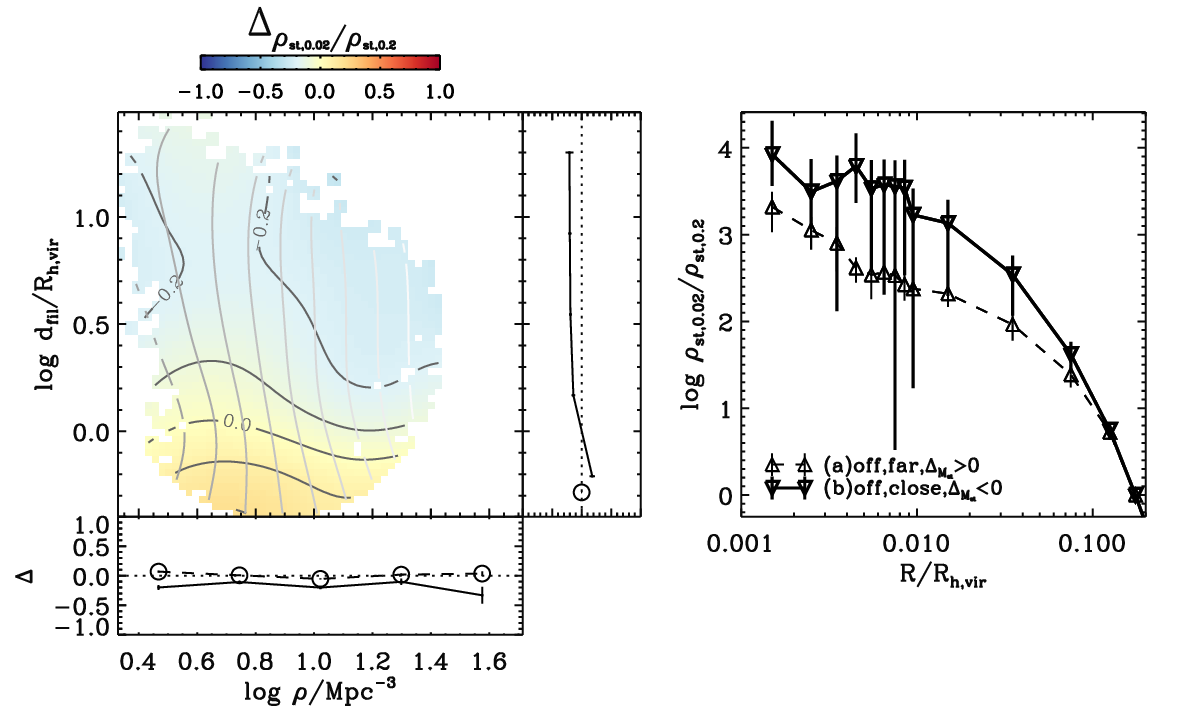}
   \caption{\textit{Left:} the residual of stellar density at 0.02$\times R_{\rm h,vir}$ divided by that at 0.2$\times R_{\rm h,vir}$, i.e., the compactness of stellar distribution ($\rho_{{\rm st}, 0.02}/\rho_{{\rm st}, 0.2}$), of ``off-filament" galaxies in the plane of distance to filaments and local density.
   Grayscale contours denote different strengths of filaments, probed by $M_{\rm sad}$, the dark matter mass around the saddle.
   Two small sub-panels on the \textit{right} and \textit{bottom} present the mean stellar compactness residuals as functions of distance to filaments and local density for the ``on-" (\textit{open circles with dashed lines}) and ``off-filament" (\textit{solid lines}) populations, respectively.
   The errors are computed from bootstrap resamplings.
   The error bars for ``on-filament" are very small.
   \textit{Right:} Stellar density profiles as a function of distance to the halo center for samples of (a) 27 galaxies far from filaments (around $d_{\rm char}$) with positive stellar mass residuals (\textit{dashed lines with triangles}) and of (b) 20 galaxies close to filaments (around $d_{\rm cut}$) with negative stellar mass residuals (\textit{thick solid lines with inverted triangles}) .
   The errors are computed from bootstrap resamplings. 
   The stellar distribution becomes more compact as galaxies are closer to filaments.
   Below 0.02$\times R_{\rm h,vir}$, error bars become much larger because the stellar distribution is not correctly resolved.}
\label{fig:AGN_z2_OffFill_rhodfil_rhost}
\end{figure*}

\begin{figure*}
   \centering
   \includegraphics[width=\textwidth]{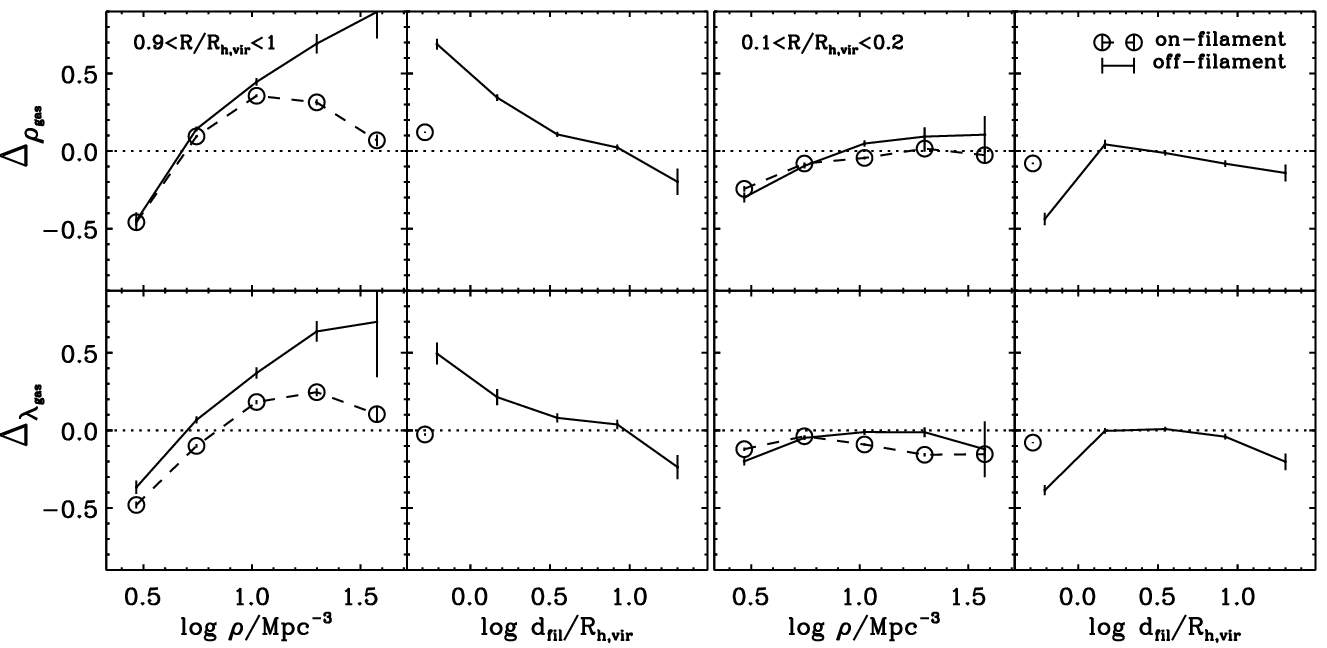}
   \caption{Mean residuals of the gas density (\textit{top}), and gas spin parameter within $0.9<R/R_{\rm h,vir}<1$ (\textit{two left columns}) and $0.1<R/R_{\rm h,vir}<0.2$ (\textit{two right columns}) as functions of local density and distance to filaments. ``Off-filament" galaxies are displayed with \textit{solid lines} and ``on-filament" galaxies with \textit{open circles with dashed lines}).
   The errors are computed from bootstrap resamplings.
   The error bars for ``on-filament" are very small.}
\label{fig:AGN_z2_1DMap_res_2}
\end{figure*}

\subsection{Gas kinematics in and around galaxies and morphology of the stellar distribution}\label{Sec:gas}
The star-formation activity in a galaxy being to first order directly dependent on the availability of a gas reservoir, we naturally turn to measure the density and kinematics of the gas in the CGM, i.e., in the host dark matter halo, in order to understand quenching in the vicinity of the filament. 

\paragraph*{The ``off-filament" population}
The five \textit{top} panels of Fig.~\ref{fig:AGN_z2_OffFil_rhodfil_rhogas_jgas} display the gas density residual (i.e., excess or deficit of gas density with respect to the gas density that is expected given the host halo mass), computed in the same way as stellar mass and SFR residual (see Section~\ref{Subsec:residual}), in shells of thickness 0.1$\times R_{\rm h,vir}$ with radii varying from 0.1$\times R_{\rm h,vir}$ to 1$\times R_{\rm h,vir}$, centered around the galaxy.
Outer shells (\textit{top left} panels) are more and more gas rich (increasing $\Delta_{\rho_{\rm gas}}$) when density increases and distance to filaments decreases, which unsurprisingly reflects that overdensities and filaments are the gas reservoirs of the Universe. 
However, at smaller shell radii (\textit{top right} panels), the trend reverts: around (shell radii $\sim0.4 R_{\rm h,vir}$) and within (shell radii $<0.2 R_{\rm h,vir}$) galaxies, we measure a lower gas density residual when the galaxies are closer to a filament than further away. 
In this sense, the gas density residual contours in the \textit{top rightmost} panel of Fig.~\ref{fig:AGN_z2_OffFil_rhodfil_rhogas_jgas} are very similar to the stellar mass or SFR residual contours in Fig.~\ref{fig:AGN_z2_OffFil_rhodfil_Mst_SFR}. 
Fig.~\ref{fig:correlation} in Appendix~\ref{App:correlations} also shows the positive correlation between stellar mass or SFR residual and gas density residual within a shell at $0.2\times R_{\rm h,vir}$, and the anticorrelation between mass or SFR residual and gas density residual within a shell at $R_{\rm h,vir}$. 
The quenching of mass assembly close to a filament is therefore related to a suppression of gas in the direct vicinity of the galaxy, not in the the vicinity of its host halo. 

In fact, the gas density residual as a function of shell radius behaves like a flattening of the CGM gas density profile, at fixed halo mass, when the halo approaches filaments.
For visualising the trend more clearly, the \textit{left} panel of Fig.~\ref{fig:profile} presents, for a sub-sample of haloes at fixed mass ($11.3<\log(M_{\rm h}/{\rm M}_{\odot})<11.6$)  and local density ($1.0< {\rm log} \rho/{\rm Mpc}^{-3} <1.5$), both the halo mass and density ranges being arbitrarily chosen, the mean gas density profiles (as a function of distance from the halo center) for (a) haloes in this subsample which sit around $d_{\rm char}$ from filaments (\textit{dashed lines with triangles}) and with positive stellar mass residuals and (b) those in the direct proximity of the filament spine around $d_{\rm cut}$ (\textit{thick solid lines with inverted triangles}) and with negative stellar mass residuals. 
This figure highlights that the slope of the gas density profile is shallower for halos in the subsample (b) in comparison to those in (a).
This flattening of the CGM gas density profile near filaments could be a natural consequence of high angular momentum of the gas component; gas particles with higher angular momentum would stay in larger orbits, leading to a shallower gas density profile \citep[see e.g.,][for a discussion in the context of dwarf galaxies]{delpopolo12}.

In order to confirm this interpretation, we compute the residual of the gas spin parameter ($\Delta_{\rm \lambda_{\rm gas}}$) defined in Eq.~\eqref{eqn:spinparam} in the same shells as those for the gas density residual.
The result is displayed in the five \textit{bottom} panels of Fig.~\ref{fig:AGN_z2_OffFil_rhodfil_rhogas_jgas}.
As expected, the spin parameter in the outer shells (\textit{bottom left} panels) is larger for haloes closer to filaments. 
This increment is in line with the theory of spin acquisition in the cosmic web through secondary infall \citep[e.g.][]{pichonetal11,laigle2015,danovichetal15}, the gas spin parameter in the outer shells naturally reflecting the angular momentum content of cosmic filaments.
This is also illustrated in the \textit{right} panel of Fig.~\ref{fig:profile}, which displays the gas spin parameter as a function of distance from the halo center for the same halo samples as used in the \textit{left} panel. 

Interestingly, we also found a decrease of gas spin in the core of haloes (\textit{bottom rightmost} panels of Fig.~\ref{fig:AGN_z2_OffFil_rhodfil_rhogas_jgas}) when their distance to filaments decreases. 
Such decrease of the spin parameter in the core of haloes seems associated with stalled gas and angular momentum transfer from outer to inner regions of the haloes (due to the large spin parameter in outer haloes).
It is also connected to increased gas dispersion,
which could be the cause of a compaction of the stellar distribution as suggested in the scenario of wet compaction \citep{Zolotovetal2015,tacchella16}.

We therefore finally measure the compactness of the stellar distribution as a function of environment.
The stellar compactness is estimated from the stellar density in a sphere of 0.02$\times R_{\rm h,vir}$ around the galaxy center\footnote{We cannot probe smaller scales than this because of the {\sc Horizon-AGN} mass resolution limit. We checked that nearly all of our sample galaxies have at least 10 stellar particles within this small sphere. The number of such galaxies drops dramatically for smaller spheres.}, divided by the stellar density at the edge of the galaxy ($\sim0.2\times R_{\rm h,vir}$).
The \textit{left} panel of Fig.~\ref{fig:AGN_z2_OffFill_rhodfil_rhost} displays the stellar compactness residual ($\Delta_{\rho_{\rm st,0.02}/\rho_{\rm st,0.2}}$) as functions of local density and distance to filaments.
While there is almost no dependency of stellar compactness residual on local density, we found that galaxies at smaller distances to filaments than $d_{\rm char}$ tend to be more concentrated than those further away, which is also illustrated in the \textit{right} panel of Fig.~\ref{fig:AGN_z2_OffFill_rhodfil_rhost} for the same subsamples as used in Fig.~\ref{fig:profile}
\footnote{Please note the large error bars for the stellar density profile of the close sample in the \textit{right} panel of Fig.~\ref{fig:AGN_z2_OffFill_rhodfil_rhost}, which is inevitable due to the small statistics used to build the profile and possibly large scatters in the sample. Nevertheless, the tendency of having a more compact stellar distribution closer to filaments is clearly visible in the mean trend in the \text{left sub-panel} of the \textit{left} panel.}.
Such compaction or bulge formation of the stellar distribution might happen prior to or simultaneously with the quenching of star formation and the flattening of the gas density profile in the CGM.
This will be further discussed in the next section.

\paragraph*{The ``on-filament" population}
We compare gas density residuals of ``off-" and ``on-filament" populations together at two radii ($R_{\rm h,vir}$ and $0.2R_{\rm h,vir}$) as functions of local density and distance to filaments in Fig.~\ref{fig:AGN_z2_1DMap_res_2}.
Unlike ``off-filament" ones, ``on-filament" galaxies do not present significant flattening of the gas density profiles in the outer halo, as inferred from the absence of the huge difference in the gas density residuals in inner and outer shells (compare the values of the open circles in the \textit{top second} panel and the \textit{top fourth} panel in Fig.~\ref{fig:AGN_z2_1DMap_res_2}).
Consistently, ``on-filament" galaxies do not show significant excess (\textit{in outer shells}) or deficit (\textit{in inner shells}) of angular momentum at a given halo mass (see the open circles in the \textit{bottom second} and \textit{fourth panels}).
These are consistent with the fact that ``on-filament" galaxies are not quenched as seen in Section \ref{Sec:SFOnVSOff}.
From the plots of gas density residual and spin parameter residual in outer haloes as a function of local density (the \textit{first} column of Fig.~\ref{fig:AGN_z2_1DMap_res_2}), ``off-" and ``on-filament" galaxies have more gas and higher angular momentum in their outer halo as local density increases (their hosting filaments being denser).
However, toward even higher densities, ``on-filament" galaxies turn out to have less gas and lower angular momentum in the outer shells.
To understand the trends (of gas densities, spin parameters as well as SFR of ``on-filament" galaxies) as a function of density, we might need to consider other factors, for example, temperature.
The compactness of the stellar distribution of ``on-filament" galaxies is on average higher than that of ``off-filament" galaxies (see the \textit{bottom sub-panel} of Fig.~\ref{fig:AGN_z2_OffFill_rhodfil_rhost}).
The \textit{right sub-panel} shows that the stellar compactness gradually increases as haloes approach filaments, peaks at the closest distance, and then decreases slightly for ``on-filament" galaxies.

\subsection{Interpretation of quenching at the edge of filaments}
The most striking result of our study is that galaxy mass assembly is less efficient precisely where the environment is gas-rich, i.e., near filaments (e.g., see the \textit{top left} panels of Fig.~\ref{fig:AGN_z2_OffFil_rhodfil_rhogas_jgas}).
We emphasise that quenching of star formation in the vicinity of cosmic filaments is specific to filaments, but is more subtle than being \textit{in} or \textit{out} filaments: this quenching mechanism is specific to the edge of cosmic filament.
Let us now sketch the broad lines of a two-step scenario in order to interpret galaxy growth and quenching at the edge of filaments.

\paragraph*{Strong inflow through gas tendrils and enhanced star formation at $d_{\rm fil}>d_{\rm char}$}
``Off-filament" haloes are connected to multiple filamentary (cold) gas tendrils at high redshifts.
As illustrated on Fig.~\ref{fig:skel2}, these gas tendrils \citep[potentially primordial, see][]{aragoncalvo16} are mainly laminar and vorticity-poor.
In addition, being multiple, they can cause counter-rotating streams, which will cancel the remaining angular-momentum and drive gas inwards towards the center \citep{algorryetal2014,danovichetal15}.
The gas inflow could be potentially clumpy, and thus dynamical friction acting on in-spiralling clumps will also help losing angular momentum and transfer gas to the center.
As haloes approach their neighbouring cosmic filaments, they are fed by more and more of these gas tendrils, and so are their central galaxies. 
Because this inflowing gas is fuel for star formation, galaxies tend to be more star-forming as they get closer to filaments, down to $d_{\rm char}$.

\paragraph*{Gas inflow stalled by vorticity-rich filaments and star formation quenched at $d_{\rm fil}<d_{\rm char}$}
When a halo approaches closer than $d_{\rm char}$ the cosmic filament, it penetrates the vorticity-rich region of the filament, at the edge of the shell-crossing region \citep{pichon99}. 
As detailed in \cite{laigle2015}, the edge of the filament is the place of highest vorticity (see also Ramsoy et al. 2020 submitted). 
Haloes in this region will therefore accrete material, and gas in particular, with coherent large angular momentum.
This is clearly illustrated in the \textit{left panels} of Fig.~\ref{fig:skel}.
The angular momentum-rich inflowing matter (a coherent fast flow) forms a large gas reservoir in the outer part of these haloes, which is described as the flattening of CGM in Section \ref{Sec:gas}.
Large angular momentum results in inefficient gas transfer to the inner halo, as discussed in e.g., \cite{pengRenzini2020,renzini2020}, starving and quenching the central galaxy.
The dependency of $d_{\rm char}$ on the halo virial radius naturally arises from the fact that we studied  galaxy properties as a function of the \textit{relative}, not absolute, distance to filaments (i.e., divided by the virial radius of each galaxy's host halo).
At first order, one could speculate that this dependency reflects in fact that larger halos live in more prominent filaments, $d_{\rm char}$ being typically set by the width of the filament defined from the vorticity truncation radius (Ramsoy et al. 2020 submitted).
Would we have more statistics, one could measure $d_{\rm char}(M_{\rm vir},M_{\rm sad})$. 
At second order indeed, the relative size of the halo with respect to the extent of the vorticity-rich region should matter. 

In addition, the stellar distribution appears to be more compact at very close distances to filaments (i.e., $d_{\rm fil}<d_{\rm char}$, see the left panel of Fig.~\ref{fig:AGN_z2_OffFill_rhodfil_rhost}).  
Previous works based on higher resolution hydrodynamical simulations \citep[e.g.][]{Martigetal2009} suggested that the presence of such a bulge is an additional way of stabilizing gas distributions.
In the so-called ``morphological quenching" scenario, the stellar bulge modifies the shape of the gravitational potential, enhancing the stability of the gas disk \citep[e.g.,][]{Gensioretal2020}.
Although in {\sc Horizon-AGN} the disc and central bulge are not resolved in their full glory due to the limited 1 kpc resolution, we can postulate that the compaction of the central galaxy, by globally modifying the shape of the gravitational potential in the halo, helps preventing the angular-momentum rich gas accreted in the outer part of the halo to be efficiently transported inward to the galaxy.
 
In conclusion, we suggest a natural mechanism for quenching star formation, which is directly related to the vorticity content of the cosmic web filaments and enhanced by the morphological quenching mechanism.

\paragraph*{What happens to the ``on-filament" population?}
The ``on-filament" galaxies do not  sit in the high-vorticity regions which are at the edge of the filaments, but in the center of those filaments \citep[where theoretically the vorticity is null, see][]{pichon99}.
Therefore, their outer haloes are not fed by the vorticity-rich gas flows, and thus the outer halo gas is more easily transferred to the inner halo, being fuel for star formation.
On the other hand, we can speculate that, as they live on average in more massive haloes, they capture more easily flows with opposite vorticity \citep[in the four-quadrant vorticity picture developed in][]{pichon99,laigle2015}, i.e., counter-rotating streams.
It is reasonable to think that this efficient angular-momentum cancellation leads to high stellar compactness, which could partly explain the definite morphological distinction between ``off-" and ``on-filament" galaxies (see the small sub-panel on the \textit{bottom} of the \textit{left} panel in Fig.~\ref{fig:AGN_z2_OffFill_rhodfil_rhost}).
We want to note that even though ``on-filament" galaxies have higher stellar compactness, they are still star-forming thanks to the angular-momentum cancellation and continuous gas feeding from outer haloes.

\section{Discussion}
\label{Sec:Discussion}

\subsection{Comparison with results from the literature}
Consistently with previous theoretical works \citep{Mussoetal2017,kraljic19a} and observational findings at low redshift \citep[e.g.][]{alpaslanetal2016,Malavasi2017,chenetal2017, poudeletal2017, laigle18,kraljic18,Sarronetal2019,bonjeanetal2019}, we confirmed that galaxies are more massive in filaments, a direct consequence that their host halo mass is larger in filaments. 
We also confirm that by $z=2$, galaxy mass assembly is less efficient in the direct vicinity of filaments than further away \textit{at fixed halo mass}. 
\cite{aragoncalvo16} assumed that the increased disorder in cosmic filaments due to several shell-crossings, or, alternatively, tidally induced shear \citep{borzyskowski17}, lets galaxies detach from their primordial gas tendrils which were previously efficiently feeding them outside the cosmic filaments, and is the reason for galaxy harassment and quenching by starvation.
In contrast, we adopt a diametrically opposed interpretation, which is supported by our finding that haloes closer to filaments have a larger gas and specific angular momentum content in their outer shells.
We highlighted a quenching mechanism driven by the kinematics of the large-scale flow in filament.
The shell-crossing of matter makes cosmic filaments vorticity-rich.
Matter accreted as secondary infall spins up both the dark matter and the gas haloes \citep[e.g.][]{pichonetal11,codisetal2012,duboisetal14,kraljic19b}.
At odds with \cite{aragoncalvo16}, we postulate that coherent motion, instead of disorder, is key for quenching galaxies.
We speculate that the increase of specific angular momentum in the outer part of the halo plays a pivotal role in hindering gas transfer to the inner halo and quenching the central galaxy, even if this halo itself keeps accreting gas.
We conclude therefore that the specificity of cosmic filaments, with their high vorticity content, cannot be ignored when trying to understand galaxy mass assembly. 

Numerous aspects of our scenario still have to be explored in more details. 
In particular we have adopted an Eulerian approach, i.e., considering the global trend for the whole galaxy population at a fixed redshift.
This approach, together with the limited resolution of {\sc Horizon-AGN} (see the discussion below in Section~\ref{Sec:caveats}) does not allow to firmly establish the full chronology of the events for galaxies flowing on filaments.
For example, it is not clear if all ``on-filament" galaxies were previously part of the ``off-filament" population.
In fact, it is likely that at least a fraction of them have formed directly on the filaments, in which case their assembly history might be different.
In the same vein, the exact processes happening to the multiple cold gas tendrils (initially connecting to the galaxy) when the galaxy travels from walls to the edge of a cosmic filament, and then finally fall at the center of this filament, are unclear.
A Lagrangian analysis, i.e., following each galaxy along its timeline, is needed to fully understand the connection between the large-scale filamentary flow and the accretion history at small-scale onto the galaxy itself, and to quantify the timing of the scenario sketched above, but this is well beyond the scope of this paper.

\subsection{Potential caveats of our study}
\label{Sec:caveats}
Although our results go in the same direction as observations at low redshift showing an excess of passive galaxies in filaments, we would like to raise several potential caveats.

\paragraph*{Statistics} The analysis must indeed first be extended to low redshift.
In addition, our analysis inevitably suffers from low statistics.
To fully disentangle the respective impact of halo mass, local density, galaxy interactions and proximity to cosmic walls/filaments on galaxy mass assembly and spin acquisition, the initial catalogues need to be divided into subsamples probing each degree of freedom, which is challenging with the size of existing simulated volumes.
Larger simulations \textit{at the same or better resolution} than {\sc Horizon-AGN} are required \citep[e.g., Horizon Run5,][]{Leeetal2020}.

\paragraph*{Filament finder algorithm}
There is no ideal tool to extract cosmic filaments \citep{libeskind18}. In this work we made the choice to extract the position of cosmic filaments from the dark matter density field using the {\sc DisPerSE} algorithm. {\sc DisPerSE} is an optimal tool as it accurately identifies the spine of the filaments (from a theoretically motivated definition) which is required to measure distance of galaxies to filaments, unlike some Hessian-based tools  which instead identify those regions of the field which are filamentary-like \citep[e.g. T-web or NEXUS+ among other, see][respectively]{foreroromero19,cautunetal2013}. However, as it was applied on the dark matter particle distribution, the position of filaments is inevitably impacted by particle noise which limits the positioning accuracy of the filaments, possibly inducing a blurring of the measured signal. Furthermore, we should keep in mind that the cut in persistence that we applied to select only the most prominent filaments implies that we potentially miss some structures in the filamentary network. Ideally, one would like to make a similar study by sorting halos depending on the persistence of the filaments they are close to \citep[e.g., as done to some extent in][]{katz2020}. Such classification is however not possible in this work since the sample is statistically limited.

\paragraph*{Resolution} More importantly, one should keep in mind the issue of the resolution limit of the simulation for the proposed scenario.
Specifically, one could argue that (a) our measurement of the compaction of the stellar distribution is not robust due to limited resolution in galaxies; (b) the stability of the gas distribution within haloes, potentially enhanced by the compact stellar distribution, might be over-predicted due to limited resolution within the galaxy and CGM (non-resolved clumps, instabilities, etc.); (c) gas temperature, that we did not discuss at all, might play an important role in quenching galaxies in filaments.

Concerning (a), using high resolution hydrodynamical simulations, \cite{duboisetal2012} investigated the formation of very compact bulges with specific angular momentum a factor of 5--30 smaller than the average angular momentum in the whole halo. 
They suggested that this bulge formation arises from effective angular momentum cancellation at the center of the halo.
In a similar vein, \cite{Zolotovetal2015} presented the compaction phase as a characteristic evolution pattern of high-redshift stream-fed galaxies \citep[see also e.g.,][]{tacchella16,wu2020}, triggered by an intense, potentially clumpy, inflow episode \citep[leading to violent disk instabilities e.g.][]{dekel14}.
This is an efficient way of quickly growing a compact bulge while turning into stars all the gas in the center of the halo. 
Their result comfort us that our scenario could still be valid with higher resolution.

Concerning (b), the fact that higher resolution simulations \citep[e.g.,][]{Gensioretal2020} precisely highlight morphological quenching argues in favour for our scenario.
Following up on these findings, \cite{dekel20} also showed that, after the compaction event, the presence of a massive bulge is key to stabilise 
a robust gaseous disk, that can eventually develop in a star-forming ring.
These features are to be explored using higher resolution simulations \citep[e.g., Extreme-Horizon][]{chabanieretal20}.

Concerning (c), the coarse resolution reached by {\sc Horizon-AGN} (like any current cosmological simulations) in the intergalactic medium (IGM) might also lead to an imperfect modelling of gas temperature.
Increased resolution in the IGM could have important consequences on galaxy properties \citep[see e.g.][]{chabanieretal20}.
Therefore our analysis could miss some important quenching mechanisms driven by the thermodynamics of the filamentary gas, simply because the resolution in the IGM does not allow to correctly model gas properties.
We therefore plan to carry out the same analysis on a higher resolution simulation. 

\paragraph*{The role of AGN feedback}
\citet{dimatteoetal12,duboisetal12agnmodel} postulated that efficient angular momentum cancellation of the gas streams, while growing the bulge, is also a way to actively grow the supermassive black hole at their center \cite[see also e.g.,][]{bournaudetal11,dimatteoetal17,dekel19,Nogueira2019}.
As a result, one could expect that AGN feedback will be boosted following the critical step of compact bulge formation.
By preventing further collapse of the high angular momentum gas in the disk, one could speculate that AGN feedback helps quenching galaxy mass assembly close to filaments.
Using {\sc Horizon-noAGN} \citep{peirani2017}, the twin simulation without AGN feedback, we will investigate in a future work how the trend changes in the absence of AGN feedback.

\section{Summary and conclusions}
\label{Sec:summary}
In this paper, we have focused on quantifying the respective impact of halo mass, local density and proximity to filaments in shaping galaxy mass and SFR at $z\sim2$.
Our exploration led us to investigate how galaxy quenching near filaments occurs in relation to the kinematics of the accreted gas and the compaction of the stellar distribution. We explicitly left aside and for a future work the topic of galaxy transformation at the nodes  of the cosmic web. 
Our main conclusions are:
\begin{itemize}
\item We extracted very precisely the geometrical location of cosmic filaments based on the dark matter particle distribution. 
The distribution of haloes around filaments is intrinsically bimodal.
We divide the sample of haloes and the galaxies that they contain into ``on-" and ``off-filament" populations.

\item Halo mass increases both as functions of local density and the inverse of distance to filaments, a trend in agreement with theoretical expectations.
Stellar mass displays a similar dependency overall, as expected from the known relation between stellar and halo mass.
However, after having controlled for halo mass, we found that galaxy mass assembly, as traced both by stellar mass and SFR, becomes more efficient down to a characteristic distance $d_{\rm char}$ from cosmic filaments and then proceeds less efficiently when further approaching a filament.
This finding is robust against the definitions of halo mass and local density, and persists at all filament strengths.
It suggests a quenching of galaxy mass assembly specific to the vicinity of the filament's spine.
Galaxies exactly ``on-filaments" are not concerned by this quenching.

\item
Outer shells of a halo contain more gas when the halo is closer to the filament spine than further away, but its galaxy itself is less gas rich.
In other words, the accreted gas onto the halo is not efficiently transferred down to the galaxy.
We postulate that this inefficient gas transfer is due to the coherent and large angular momentum of the outer halo gas fed by vorticity-rich filaments.
The gas transfer could be made even less efficient by the presence of the central stellar bulge (through further stabilization of the distribution of outer halo gas).

\item Altogether and in spite of well identified numerical limitations, our results suggest that the specificity of the filamentary environment, with its high vorticity content, cannot be ignored when trying to understand galaxy mass assembly and quenching. 
\end{itemize}

We emphasise that interpreting the redshift evolution of the trend is not straightforward, as we do not know \textit{a priori} how a galaxy population at a given redshift evolved into a population at a latter redshift.
In particular, it is not obvious that the quenching mechanism highlighted in our study can fully explain the low redshift observations (more passive galaxies closer to filaments than star-forming ones).
Extending our analysis to a Lagrangian approach and lower redshift will be the topic of a future work.

\vskip 0.1cm
\section*{Acknowledgments}
{\sl 
This research was supported by Basic Science Research Program through the National Research Foundation of Korea(NRF) funded by the Ministry of Education (2020R1I1A1A01069228).
This work was granted access to the HPC resources of CINES (Jade) under the allocation 2013047012 and c2014047012 made by GENCI.
This work was supported by the Programme National Cosmology and Galaxies (PNCG) of CNRS/INSU with INP and IN2P3, co-funded by CEA and CNES, and by ANR spin(e) ANR-13-BS05-0005. Let us thank D.~Munro for freely distributing his {\sc \small  Yorick} programming language and opengl interface (available at \url{http://yorick.sourceforge.net/}).  
We  warmly thank S.~Rouberol for running  the {\tt Horizon} cluster on which the simulation was  post-processed. 
The light cone of horizon-AGN is visible at \url{http://skymaps.horizon-simulation.org}.
CL would like to thank Stephane Arnouts for stimulating discussions and the Korean Astronomy and Space science Institute for hospitality when part of this work was done.
HS would like to thank Taysun Kimm for helpful advices and discussions.
}

\section*{Data availability}
The data underlying this article were accessed from \url{http://www.horizon-simulation.org/}. The derived data generated in this research will be shared on reasonable request to the corresponding author.

\bibliographystyle{mn2e}
\bibliography{ms}

\appendix
\section{Alternative measures for local density}
\label{Sec:alter}
So far we have considered the effect of relatively large-scale environment, i.e., local density smoothed over 1~Mpc and distance to cosmic filaments of high persistence.
However, the effect of small-scale environment should be also taken into account because (1) small-scale environment, i.e., proximity to neighbour galaxies, is also one of the most important factors that can alter mass assembly \citep[e.g.,][]{Parkchangbom09,sabater15}, and (2) small-scale environment is correlated with large-scale environment \citep[see e.g., a study of the impact of scales on measuring density in][]{kraljic18}.
Therefore, it is important to check that the scale at which density is estimated does not alter our conclusions on the role of cosmic filaments in shaping mass assembly of galaxies.

\begin{figure*}
   \centering
   \includegraphics[scale=0.8]{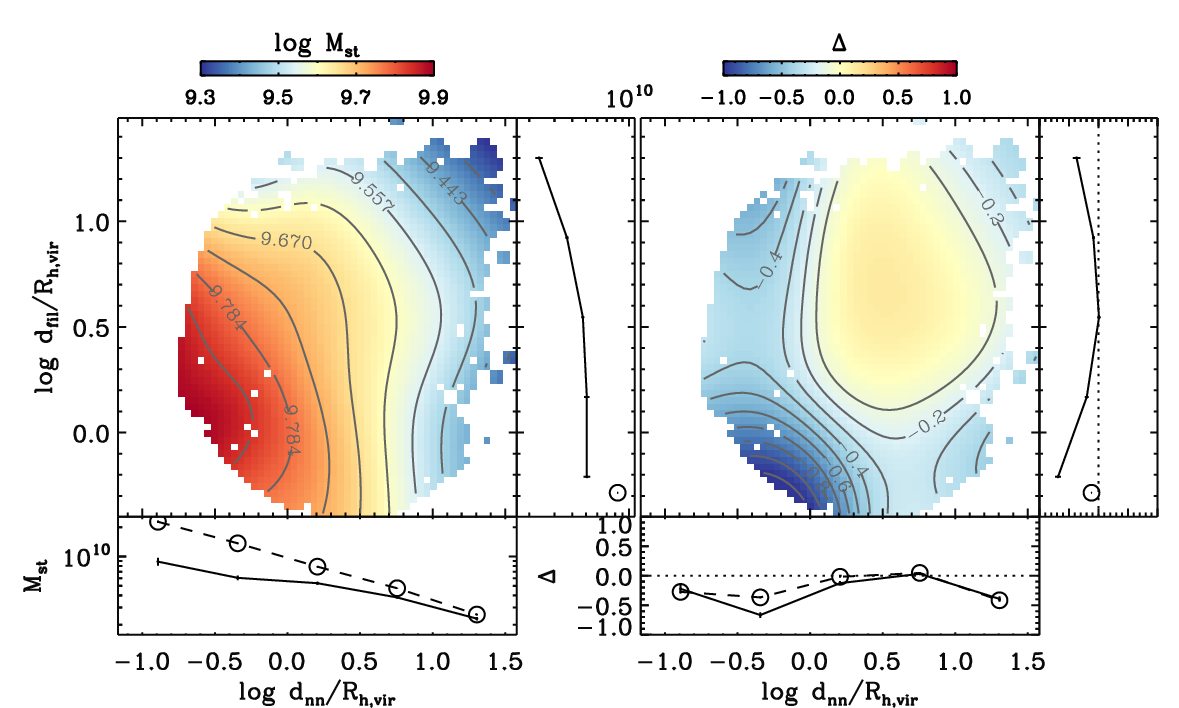}
        \caption{Similar to the top panels of Fig.~\ref{fig:AGN_z2_OffFil_rhodfil_Mst_SFR}, but using for the $x$-axis the distance to the nearest neighbouring galaxy (divided by the halo virial radius) instead of the density smoothed over 1~Mpc.
        In this plane, the trend of stellar mass residual towards filaments, at fixed distance to the nearest neighbour, is preserved.
        In the subpanels, mean values as functions of distance to the nearest neighbor (\textit{bottom}) and of distance to filaments (\textit{right}) are presented for ``off-" (\textit{solid lines} with error bars) and ``on-filament" (\textit{open cicles with dashed lines}) populations.
        The errors are computed from bootstrap resamplings.}
\label{fig:AGN_z2_OffFil_dnndfil_Mst}
\end{figure*}

\begin{figure*}
   \centering
   \includegraphics[scale=0.8]{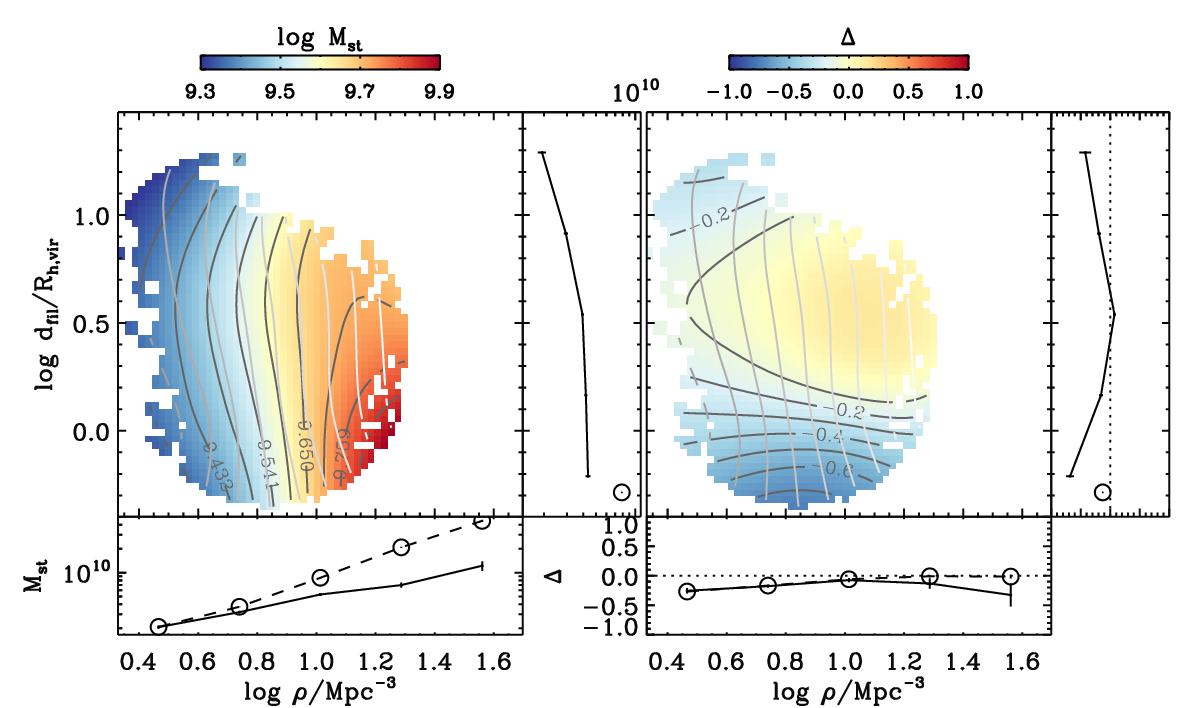}
        \caption{Same as the top panels of Fig.~\ref{fig:AGN_z2_OffFil_rhodfil_Mst_SFR}, but after having removed galaxies which are closer to their neighbouring galaxy than to their closest cosmic filament.
        The trend of stellar mass residual towards filaments, at fixed density, is preserved.
        Grayscale contours denote different strengths of filaments, probed by $M_{\rm sad}$, the dark matter mass around the saddle.
        In the subpanels, mean values as functions of local (\textit{bottom}) and of distance to filaments (\textit{right}) are presented for ``off-" (\textit{solid lines} with error bars) and ``on-filament" (\textit{open cicles with dashed lines}) populations.
        The errors are computed from bootstrap resamplings.}
\label{fig:AGN_z2_OffFil_dfilLTdnn_rhodfil_Mst}
\end{figure*}

We measured the distance to the closest galaxy (either central or satellite; i.e., nearest neighbour) as a small-scale density probe.
We then traced stellar mass residual in a two-dimensional plane similarly as before, but replacing the density smoothed over 1~Mpc by distance to the nearest neighbour.
In Fig.~\ref{fig:AGN_z2_OffFil_dnndfil_Mst}, the contours of stellar mass and its residual on the plane of distance to the nearest neighbor and distance to filaments shows that the trend of decreasing residual at the edge of the filament is preserved whatever the density estimator.
We confirmed this in another way; we removed from the sample all the galaxies that are closer to their neighbour galaxy than their host filament, and then made the stellar mass residual map on the original plane of local density-distance to filaments (Fig.~\ref{fig:AGN_z2_OffFil_dfilLTdnn_rhodfil_Mst}).
Even though the impact of neighboring galaxies is minimized, the trend of the stellar mass residual with distance to filaments remains as seen in the top right panel of Fig.~\ref{fig:AGN_z2_OffFil_rhodfil_Mst_SFR}.
This indicates that the unique role of filaments in galaxy mass assembly, that is distinct from those of large-scale and small-scale density, does exist.

\section{Gas density, spin parameter and stellar compactness as a function of halo mass at $z\sim2$}
\begin{figure*}
   \centering
   \includegraphics[width=0.9\textwidth]{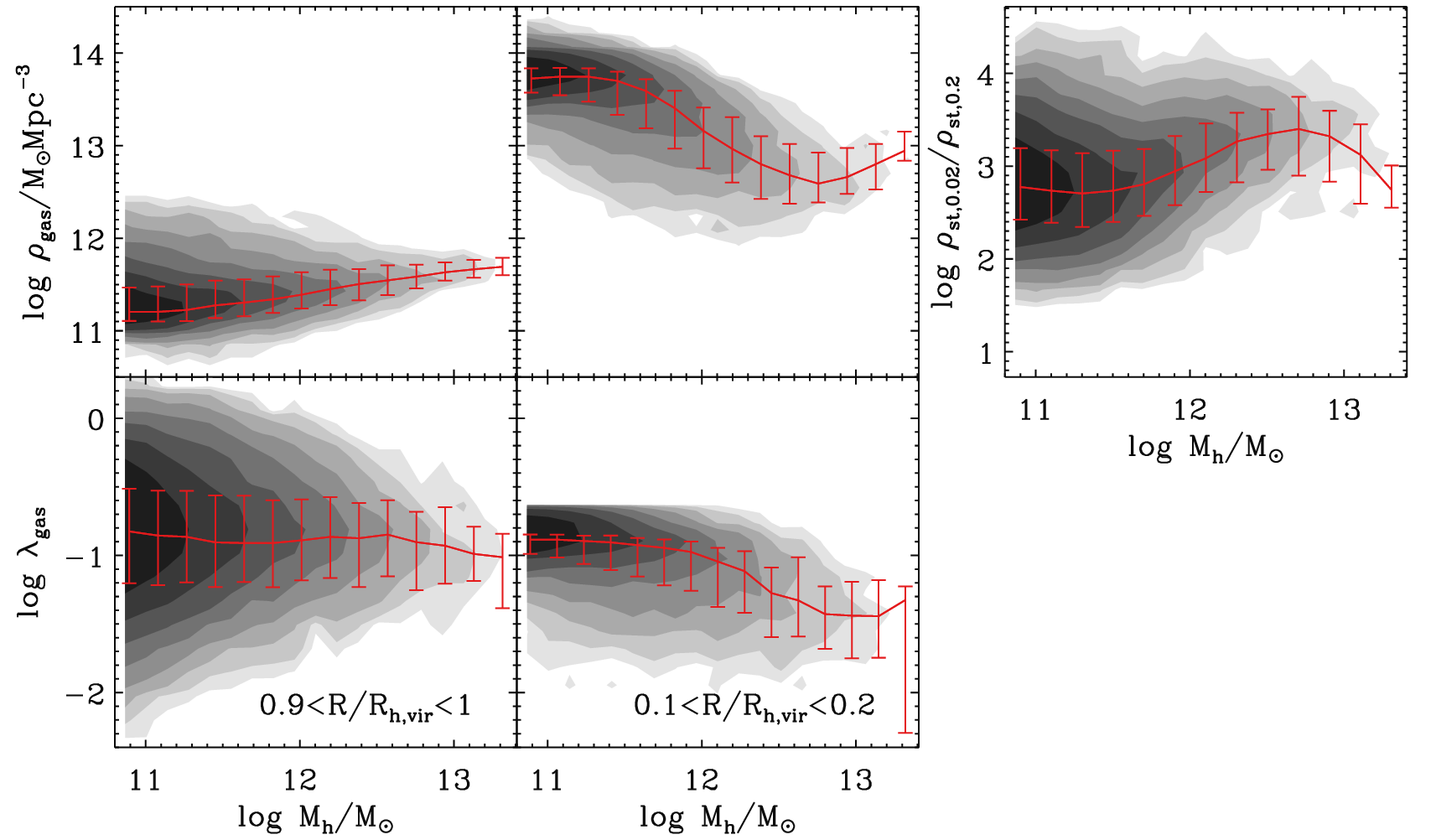}
        \caption{Gas density (\textit{top}) and spin parameter (\textit{bottom}) in shells of thickness 0.1$\times R_{\rm h,vir}$ at radii of 0.9$\times R_{\rm h,vir}$ (\textit{left}) and 0.1$\times  R_{\rm h,vir}$ (\textit{middle}), and stellar compactness ($\rho_{{\rm st}, 0.02}/\rho_{{\rm st}, 0.2}$, \textit{right}) as a function of halo mass for central galaxies in {\sc Horizon-AGN} at $z=1.97$.
        Red line with error bars in each panel are the median trend with 1$\sigma$ scatter.}
\label{fig:AGN_z2_Mh_GasRhoSpinpar}
\end{figure*}

\begin{figure*}
   \centering
   \includegraphics[width=0.8\textwidth]{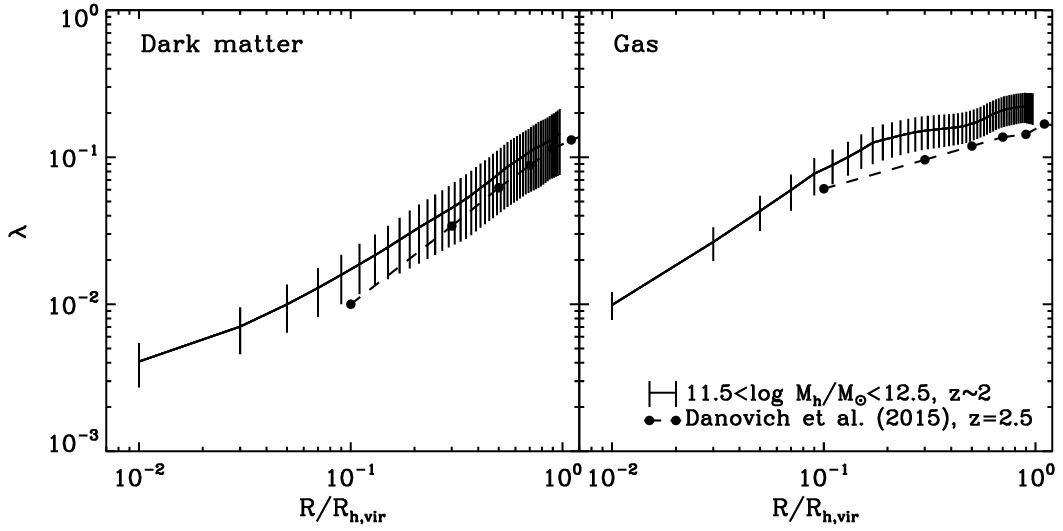}
        \caption{Radial profiles of spin parameters of dark matter (\textit{left}) and gas (\textit{right}) for central galaxies of $11.5<\log M_{\rm h,vir}/M_{\odot}<12.5$ in {\sc Horizon-AGN} at $z=1.97$ (\textit{solid line with error bars}), which are compared with those in \citet[][see their Fig.~1]{danovichetal15} for similar masses and redshift (\textit{filled circles with dashed line}).}
\label{fig:profile2}
\end{figure*}

Fig.~\ref{fig:AGN_z2_Mh_GasRhoSpinpar} shows the median trends of the gas density and spin parameter in shells of thickness $0.1\times R_{\rm h,vir}$ at two radii $0.9\times R_{\rm h,vir}$ (\textit{left}) and $0.1\times R_{\rm h,vir}$ (\textit{middle}) as a function of halo mass.
The gas density in the outer shell increases as a function of halo mass, whereas that in the inner shell decreases.
The gas spin parameter tends to slightly decreases as a function of halo mass whatever the shell radius (but more strongly in smaller shells).
This result is broadly consistent with \citet{knebe&power08}, who measured the redshift evolution of the dependency of the dark matter (total, not in shell) spin parameter on halo mass \citep[see also e.g.][]{bett07}.
As another cross-check, we compare the radial profiles of the spin parameters of dark matter and gas to those shown in \citet{danovichetal15} (for haloes of similar masses at redshifts).
Fig.~\ref{fig:profile2} shows in general agreements between ours and those of \cite{danovichetal15} with systematic slight overestimation of ours due to the resolution limit of {\sc Horizon-AGN}, as mentioned in Section~\ref{Subsec:cat}.
The stellar compactness (i.e., the stellar density ratio at $0.02\times R_{\rm h,vir}$ and $0.2\times R_{\rm h,vir}$) increases as a function of halo mass (\textit{right} panel).

\section{Correlation between residuals}
\label{App:correlations}
\begin{figure*}
   \centering
    \includegraphics[width=0.99\textwidth]{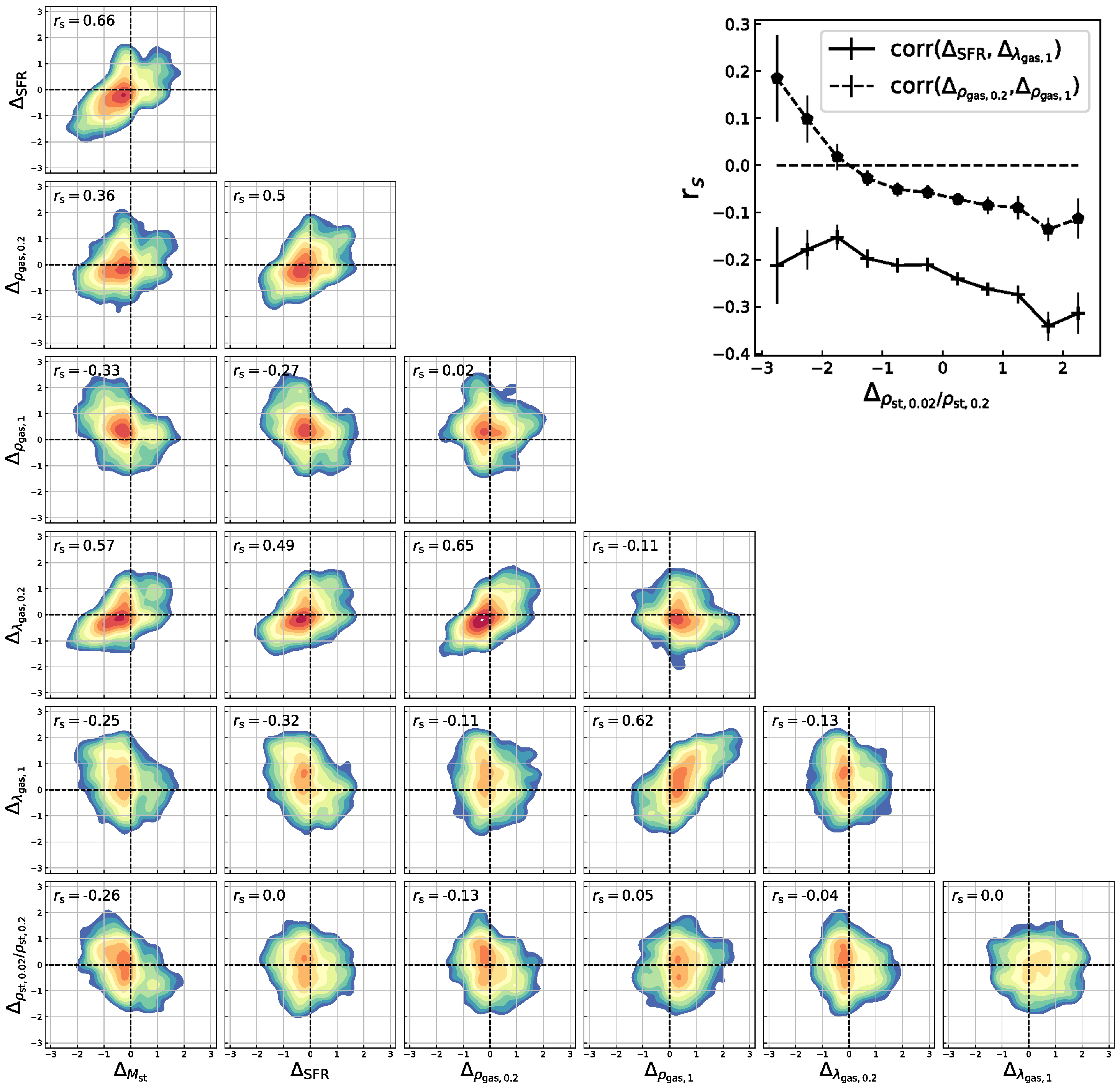}
        \caption{Correlation, together with Spearman correlation coefficients, between the residuals of stellar mass, SFR, gas mass and spin parameter at 0.2$\times$ and 1$\times R_{\rm h,vir}$, and compactness of the stellar distribution (as estimated from the density within 0.02 halo virial radius) for ``off-filaments" galaxies sitting closer than $d_{\rm char}$ to the cosmic filaments.
        The sub-panel on the \textit{top right} of the plot shows how the Spearman correlation coefficient between the residuals of SFR and gas spin parameter in the outer halo shell on the one hand (\textit{solid} line), and that between the residuals of gas density in the inner and outer shells on the other hand (\textit{dashed} line) evolves as a function of the residual of stellar distribution compactness. The errors are computed from bootstrap resamplings.}
\label{fig:correlation}
\end{figure*}

Fig.~\ref{fig:correlation} presents the correlation, together with Spearman correlation coefficients, between the residuals of stellar mass, SFR, gas density and spin parameter at 0.2$\times$ and 1$\times R_{\rm h,vir}$, and compactness of the stellar distribution for ``off-filaments" galaxies sitting closer than $d_{\rm char}$ to the cosmic filaments.

The recovered correlations are in line with the argument developed in the text: the anti-correlations of the gas density residual or spin parameter residual at $R_{\rm h,vir}$ with stellar mass residual and SFR residual, respectively.
We can note in particular the negative correlation between the residuals of the gas spin parameter at $R_{\rm h,vir}$ and of SFR (\textit{second left panel on the second bottom row}), which indicates a quenching mechanism playing at the vicinity of filament boundaries suggested in this study.

Although the Spearman coefficient is null for the correlation between the residuals of the stellar compactness and SFR (\textit{second left panel on the bottom row}), we note that the mode in the distribution of this residual pair corresponds to $\Delta_{\rm SFR}<0$ and $\Delta_{\rho_{\rm st,0.02}}>0$.
This supports the idea that a larger compactness is associated with the quenching of the galaxy.
There is also no direct correlation between the residual of the stellar compactness and the residual of the gas spin parameter in the outer shell (\textit{rightmost panel on the bottom row}).
However, we address the impact of the stellar compactness on star formation and quenching in an indirect way.
The \textit{top right sub-panel} displays the variation of Spearman correlation coefficient as a function of the stellar compactness residual for two specific correlations: $\Delta_{\rm SFR}$--$\Delta_{\lambda_{\rm gas,1}}$ and $\Delta_{\rho_{\rm gas,0.2}}$--$\Delta_{\rho_{\rm gas,1}}$.
The anti-correlation between the residuals of SFR and of the gas spin parameter in the outer shell (\textit{solid} line) becomes stronger when the stellar compaction increases.
The same phenomena happens for the anti-correlation between the gas density in the inner and outer shells (\textit{dashed} line).
This supports the idea that a larger compactness favours galaxy quenching.

\end{document}